\documentclass[twocolumn]{aastex63}
\usepackage{amsmath}
\usepackage{multirow}
\usepackage{tablefootnote}

\def \nustar {{NuSTAR}}

\def \maxi {MAXI~J1848–015}
\def \caltech {{Cahill Center for Astronomy and Astrophysics, California Institute of Technology, Pasadena, CA 91125, USA}}

\newcommand{\code}{\texttt}
\newcommand{\comment}[1]{}

\shorttitle{\maxi}
\shortauthors{Pike et al.}
\graphicspath{{./}{}}
\begin{document}

\title{MAXI and NuSTAR observations of the faint X-ray transient MAXI~J1848–015 in the GLIMPSE-C01 Cluster}

\correspondingauthor{Sean N. Pike}
\email{spike@caltech.edu}

\author[0000-0002-8403-0041]{Sean N. Pike}
\affiliation{\caltech{}}

\author[0000-0003-0939-1178]{Hitoshi Negoro}
\affiliation{Department of Physics, Nihon University, 1-8 Kanda-Surugadai, Chiyoda-ku, Tokyo 101-8308, Japan}

\author[0000-0001-5506-9855]{John A. Tomsick}
\affiliation{Space Sciences Laboratory, 7 Gauss Way, University of California, Berkeley, CA 94720-7450, USA}

\author[0000-0002-4576-9337]{Matteo Bachetti}
\affiliation{INAF–Osservatorio Astronomico di Cagliari, via della Scienza 5, I-09047 Selargius (CA), Italy}

\author[0000-0002-4024-6967]{McKinley Brumback}
\affiliation{\caltech{}}

\author[0000-0002-8908-759X]{Riley M. T. Connors}
\affiliation{\caltech{}}

\author[0000-0003-3828-2448]{Javier A. Garc\'ia}
\affiliation{\caltech{}}

\author[0000-0002-1984-2932]{Brian Grefenstette}
\affiliation{\caltech{}}

\author[0000-0002-8548-482X]{Jeremy Hare}
\affiliation{NASA Goddard Space Flight Center, Greenbelt, MD 20771, USA}

\author{Fiona A. Harrison}
\affiliation{\caltech{}}

\author[0000-0002-3850-6651]{Amruta Jaodand}
\affiliation{\caltech{}}

\author[0000-0002-8961-939X]{R.~M.~Ludlam}\thanks{NASA Einstein Fellow}
\affiliation{\caltech{}}

\author[0000-0003-4216-7936]{Guglielmo Mastroserio}
\affiliation{\caltech{}}

\author[0000-0002-6337-7943]{Tatehiro Mihara}
\affiliation{RIKEN, Hirosawa, Wako, Saitama, 351-0198, Japan}

\author[0000-0001-8195-6546]{Megumi Shidatsu}
\affiliation{Department of Physics, Ehime University, 
2-5, Bunkyocho, Matsuyama, Ehime 790-8577, Japan}

\author[0000-0002-1190-0720]{Mutsumi Sugizaki}
\affiliation{National Astronomical Observatories, Chinese Academy of Sciences, 20A Datun Road, Beijing 100012, People's Republic of China}

\author{Ryohei Takagi}
\affiliation{Department of Physics, Nihon University, 1-8 Kanda-Surugadai, Chiyoda-ku, Tokyo 101-8308, Japan}

% \nocollaboration{2}



\begin{abstract}

We present the results of MAXI monitoring and two \nustar\ observations of the recently discovered faint X-ray transient \maxi. Analysis of the MAXI light-curve shows that the source underwent a rapid flux increase beginning on 2020 December 20, followed by a rapid decrease in flux after only $\sim5$ days. \nustar\ observations reveal that the source transitioned from a bright soft state with unabsorbed, bolometric ($0.1$--$100$\,keV) flux $F=6.9 \pm 0.1 \times 10^{-10}\,\mathrm{erg\,cm^{-2}\,s^{-1}}$, to a low hard state with flux $F=2.85 \pm 0.04 \times 10^{-10}\,\mathrm{erg\,cm^{-2}\,s^{-1}}$. Given a distance of $3.3$\,kpc, inferred via association of the source with the GLIMPSE-C01 cluster, these fluxes correspond to an Eddington fraction of order $10^{-3}$ for an accreting neutron star of mass $M=1.4M_\odot$, or even lower for a more massive accretor. However, the source spectra exhibit strong relativistic reflection features, indicating the presence of an accretion disk which extends close to the accretor, for which we measure a high spin, $a=0.967\pm0.013$. In addition to a change in flux and spectral shape, we find evidence for other changes between the soft and hard states, including moderate disk truncation with the inner disk radius increasing from $R_\mathrm{in}\approx3\,R_\mathrm{g}$ to $R_\mathrm{in}\approx8\,R_\mathrm{g}$, narrow Fe emission whose centroid decreases from $6.8\pm0.1$\,keV to $6.3 \pm 0.1$\,keV, and an increase in low-frequency ($10^{-3}$--$10^{-1}$\,Hz) variability. Due to the high spin we conclude that the source is likely to be a black hole rather than a neutron star, and we discuss physical interpretations of the low apparent luminosity as well as the narrow Fe emission.

\end{abstract}


\section{Introduction}\label{sec:intro}

\subsection{X-ray binaries in outburst}
Compact objects, such as neutron stars (NS) and black holes (BH), orbiting main sequence stars often undergo cycles of outburst and quiescence due to modulation in the rate at which matter from the companion star accretes onto the compact object. Matter falls from the surface of the companion into the orbit of the compact object via Roche-lobe overflow or stellar winds (also known as Bondi-Hoyle accretion). According to the Disk Instability Model \citep{Lasota2001}, this material eventually reaches a critical density at which angular momentum can be transported efficiently outward, resulting in the formation of an accretion disk \citep{Shakura1973}. This disk may reach all the way down to the surface or inner-most stable circular orbit (ISCO) of the compact object. As the gravitational potential energy of the accreting material is converted into heat, the innermost regions of the disk can reach temperatures of order $10^7$\,K, emitting photons with energy exceeding thousands of electron volts, hence these systems are referred to as X-ray binaries. This cycle of transient disk accretion leads to a variety of observable phenomena. The X-ray spectra of accreting NSs and BHs vary between a low-luminosity hard state and a brighter soft state, with intermediate states in between. During the hard state, emission is dominated by a power-law-shaped component which originates from a region of hot plasma near the central accretor known as the corona, while the soft state spectrum is dominated by thermal emission from the inner regions of the disk.

Additional spectral features result from the disk geometry in an X-ray binary. Iron in the disk may be irradiated by coronal emission and fluoresce, giving rise to emission lines around $6.4$\,keV. Line emission originating from the inner regions of the disk may be blurred by Doppler shifts due to the rapid orbital motion of the disk material and by the strong gravitational potential near the central compact object. Furthermore, soft photons from the corona may undergo Compton upscattering resulting in a ``Compton hump" in the spectrum at high energies. These ``reflection" features encode information regarding the inner disk radius, $R_\mathrm{in}$, and the spin parameter, $a$, of the central accretor. Spectral models such as \code{relxill} \citep{Dauser2014,Garcia2014}, which self-consistently model relativistic disk reflection, are therefore important tools for probing the properties of X-ray binaries, and can help us to differentiate between NS and BH accretors.

\subsection{\maxi}
Each year, the Monitor of All-sky X-ray Image, or MAXI \citep{Matsuoka2009}, discovers dozens of X-ray sources. Among these sources are accreting black holes, neutron stars, and white dwarfs. Follow-up with other X-ray observatories can help to elucidate the nature of these sources and lays the groundwork for their future study.

One such source, \maxi, was discovered with the MAXI/GSC on December 20, 2020 \citep{Takagi2020}.
The MAXI/GSC nova alert system \citep{Negoro2016} triggered on the source at 05:04 (all times are given in UT), and 
the source flux was found to be increasing. The average X-ray flux on eight scan transits from 00:25 to 11:16 
on December 20 was $63\pm10$ mCrab in the 4--10 keV band. 
The MAXI $90\%$ confidence region had a radius of about $0.3$\,deg and was consistent with the previously detected ASCA source AX~J1848.8-0129 \citep{Sugizaki2001}.
Since the source had an angular separation of $26$\,deg from the Sun at the time of detection, neither the Neil Gehrels Swift Observatory X-ray Telescope \citep[Swift/XRT,][]{Gehrels2004,Burrows2005} nor the Neutron Interior Composition Explorer Mission \citep[NICER,][]{Gendreau2017} could observe the source. 

However, the Nuclear Spectroscopic Telescope Array, or \nustar\ \citep{Harrison2013}, was able to perform follow-up observations of the source despite its angular proximity to the Sun. \nustar\ performed tiling observations to search the MAXI error region, and the source was detected during the first pointing. The source exhibited a soft spectrum, and although the source position was further refined to a region with radius $\sim90^{\prime\prime}$, it could still not be distinguished from {AX~J1848.8-0129} with certainty \citep{Pike2020}. Soon after the \nustar\ ToO observations, on December 23, the 2--10\,keV source flux rapidly decreased in a day
from about 40 mCrab to less than 15 mCrab \citep{Negoro2020}. About a week later, the source was again observed by \nustar, this time exhibiting a much harder spectrum \citep{Mihara2021}.

By the end of February 2021, \maxi\ was far enough from the Sun that it could be observed by other instruments. Thereafter, the source was localized by both Swift/XRT and the Chandra X-ray Observatory \citep[Chandra,][]{Weisskopf2000} to $90\%$ confidence regions of $\pm2.3^{\prime\prime}$ \citep{Kennea2021} and $\pm0.8^{\prime\prime}$ \citep{2021ATel14424....1C}, respectively, while radio observations further refined the source position to $\alpha\mathrm{(J2000) = 18h\,48m\,49.824s \pm 0.003s}$, $\delta\mathrm{(J2000) = -01^\circ\, 29^\prime\, 49.99^{\prime\prime} \pm 0.05^{\prime\prime}}$ \citep{Tremou2021}.\footnote{The authors do not specify whether the uncertainties represent $1\sigma$ or $90\%$ confidence regions.} Interestingly, the radio position is coincident with a relatively bright NIR counterpart \citep{Hare2021}. Around the same time, NICER observations of the source were performed and \citet{Miller2021} reported a number of emission lines in the Fe K band and a flux of about $1$\,mCrab.

Importantly, the precise localization of the source confirmed that \maxi\ is not consistent with the previously reported position of AX~J1848.8-0129, and that it is spatially coincident with, and likely resides in, the core of the Galactic cluster GLIMPSE-C01 (GC01 hereafter). This cluster was originally discovered by \cite{Kobulnicky2005}, who suggested that the cluster was an old globular cluster passing through the Galactic disk at a distance of $3.1-5.2$\,kpc. Several subsequent studies have alternatively suggested that GC01 is a young or intermediate age massive cluster candidate with an age between $0.3-2.5$\,Gyr (\citealt{Davies2011,Davidge2016}). More recently, \cite{Hare2018} reported on Hubble Space Telescope observations of GC01, which they used to estimate a cluster distance of $\sim3.3$\,kpc and a cluster age of $>2$\,Gyr by studying the absolute magnitudes of red clump stars in the cluster. Unfortunately, the cluster’s large source density, strong differential reddening across the cluster (ranging between $A_V=14-22$), and unknown metallicity make it difficult to more precisely constrain the cluster’s age \citep{Hare2018}.

In this paper, we present the results of MAXI monitoring of \maxi\ as well as spectral and timing analysis of the two \nustar\ observations performed following the detection of the source by MAXI. In Section \ref{sec:maxi_obs}, we describe the MAXI observations in detail, including the shape and duration of the outburst. In Section \ref{sec:nustar}, we discuss the \nustar\ observations of the source, beginning with a description of the observations and data reduction and continuing onto an investigation of the source spectra of as well as the results of reflection modeling applied to these spectra in Section \ref{sec:spectra}. Next, we present an analysis of the timing properties of the \nustar\ light-curves in Section \ref{sec:nustar_timing}. Finally, in Section \ref{sec:discussion}, we discuss what our results mean in regards to the questions of whether the source is a neutron star or a black hole accretor and how its particularly low luminosity can be understood in the context of disk accretion. 

\section{MAXI Observations} \label{sec:maxi_obs}

\subsection{Observations and data reduction}

MAXI has been monitoring about 85\% of sky every 92 minutes 
with the Gas Slit Cameras, GSCs, in the 2--20 keV band since August 2009 \citep{Mihara2011,Sugizaki2011}.
The GSCs have two wide fields of view of $1.5^{\circ} \times 160^{\circ}$ to the horizontal and zenith directions,
and typically observe a source for 40--100 seconds every 92 minutes as the International Space Station (ISS) orbits Earth.

In December 2020, the GSC\_1, GSC\_2, GSC\_4, GSC\_5, and GSC\_7 cameras and the degraded GSC\_3 and GSC\_6 cameras were operating. The source was detected by all these detectors, but in these analyses we only used those obtained with the well-calibrated GSC\_4 and GSC\_5 cameras with a high voltage of 1650\,V and GSC\_2 and GSC\_7 cameras with 1550\,V.

We employed a point-spread function (PSF) fit method to obtain MAXI/GSC light-curves with the best signal-to-noise ratios \citep{Morii2010}. The count rate in each light-curve bin was obtained by fitting an image with the PSFs of the GSCs taking into account the presence of nearby sources such as the high-mass X-ray binary AX J1846.4-0258 (1.60 degree separation) and the Super-giant Fast X-ray Transient IGR J18483$-$0311 (1.68 degree separation). 
On the other hand, the nearby transient sources in quiescent states, including Swift J185003.2$-$005627 (0.64 deg), GS 1843$-$02 (0.93 deg),  and Swift J1845.7$-$0037 (1.11 deg), were ignored.
AX J1848.8$-$0129 (0.012 deg $\simeq$ 0.74 arcmin) was also excluded. We note that AX J1848.8$-$0129 is originally named as AX J184848$-$0129 \citep{Sugizaki2001}, and its position has about 1 arcmin uncertainty.
We first produced 1-scan, 6-hr, and 1-day bin light-curves 
in the 2--4 keV and 4--10 keV bands, respectively. The 6-hr and 1-day bin data were obtained by fitting 
4-scan and 16-scan image data, respectively.
We then subtracted constant background components, mainly originating from the galactic ridge X-ray emission, as the average count rates from MJD 58000 (2017 September 4) to 58599 (2019 April 26) were consistent with zero. 
Finally, we obtained 2--10 keV curves by summing the background subtracted 2--4 keV and 4--10 keV curves.

As shown later, every $\sim92$ days (the ISS precession period), when the source was at the right side of the GSC\_2 and  GSC\_5,
the source was obscured by the Space-X Crew-1 spacecraft attached on the Harmony module of the ISS for several days. We did not use data during the periods shown in the current calibration database (CALDB)
and we also excluded data from 1 day before and after this period due to some ambiguity regarding the shape of the spacecraft shadow.

\subsection{Long-term light-curve} \label{sec:maxi_flux}

\begin{figure}
\begin{center}
\includegraphics[width=0.36\textwidth,angle=90]{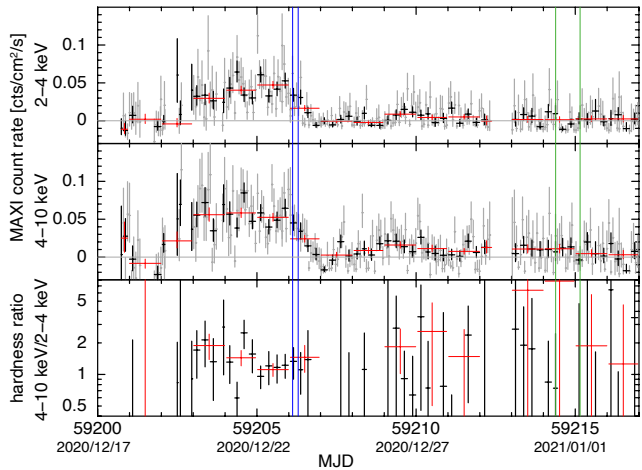}
\caption{2--4 keV and 4--10 keV light-curves (upper and middle panels) and their ratios (lower) of \maxi\ obtained with MAXI/GSC.
Average fluxes in 1 scan (grey), 6 hours (black), and 1 day (red) are shown (1 scan data are omitted in the lower panel). The average 2--4 keV and 4--10 keV count rates for Crab are 1.065 and 1.172\,ct\,cm$^{-2}$\,s$^{-1}$, respectively.
The blue and green lines are the start and end time of the first and second \nustar\, observations, respectively.}
\label{fig:maxilcshort}
\end{center}
\end{figure}

\begin{figure}
\begin{center}
\includegraphics[width=0.46\textwidth]{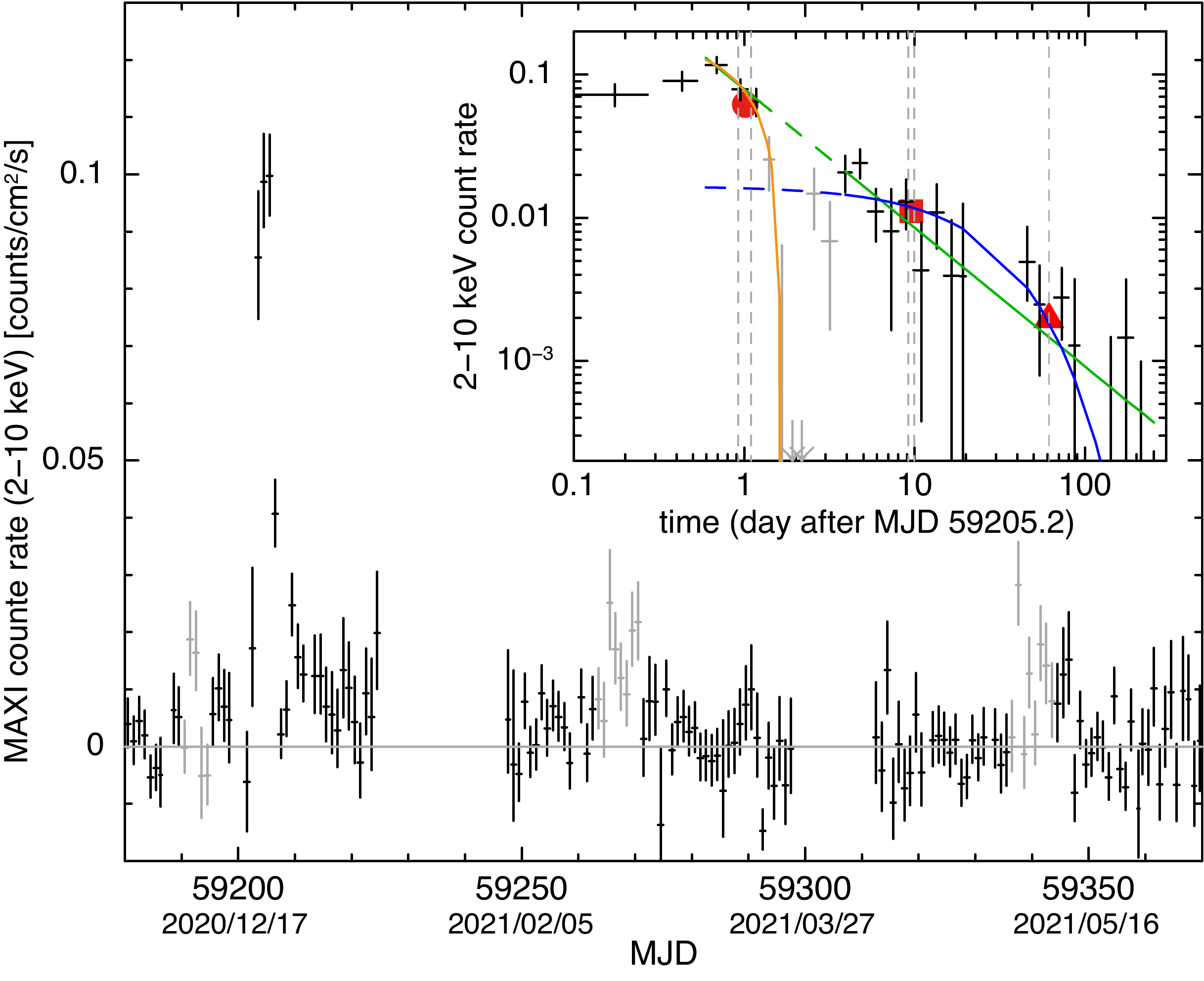}
\caption{Long-term 2--10 keV light-curve of \maxi\ from MAXI. Unused data during interference with the Space-X Crew-1 spacecraft are also shown in grey.
A logarithmically rebinned curve and the best fitting curves are also shown in the inset panel. The data during a dip shown in grey are not used in the fits (see Section \ref{sec:maxi_flux}).
Observed fluxes for the two \nustar\ and Swift observations are indicated with the red filled circle, 
rectangle, and triangle, respectively, between the dashed lines showing the observation periods. The orange line shown in the inset represents a linear fit to the initial rapid decay, while the blue and green lines represent exponential and power law fits, respectively, to the observed second decay.}
\label{fig:maxilclong}
\end{center}
\end{figure}

Figure~\ref{fig:maxilcshort} shows the MAXI/GSC 2--4 keV and 4--10 keV light-curves and their ratio for \maxi\ obtained by the PSF-fit method.
In the seven scan transits from 18:37 on December 18 to 01:12 on December 19 (MJD 59201.7763 -- 59202.0501), the source was not visible on GSC images \citep{Takagi2020}. No count excess is recognized in each scan curve. The PSF-fit method provides two-sided 1$\sigma$ errors based on the likelihood method \citep{Morii2010}. Fitting eight one-scan data points during the period with a constant model gives the average count rate of -0.094$\pm$0.0128\,ct\,cm$^{-2}$\,s$^{-1}$. Using the size of this error, we estimate a 1$\sigma$ upper limit of 0.0128\,ct\,cm$^{-2}$\,s$^{-1}$, $\sim 5.7$ mCrab\footnote{We adopt the 2--10 keV count rate and the flux for Crab as 2.237\,ct\,cm$^{-2}$\,s$^{-1}$ and $2.4 \times 10^{-8}$\,erg\,cm$^{-2}$\,s$^{-1}$, respectively.}. This is consistent with a value 12 mCrab of a typical 3$\sigma$ detection limit of 1 day of $\sim16$ scans \citep{Negoro2016}.
After the scan transit at 01:12 until 22:52 on December 19 (MJD 59202.05-59202.95),  the GSC did not fully observe the source region due to Sun avoidance.  
During the three scan transits from 11:59 to 15:05, however, some count excess was recognized at the edge of the detectors, which suggests the outburst started between 02:22 and 11:59.

The light-curves and hardness ratios in Fig. \ref{fig:maxilcshort} show that the 4--10 keV flux reached the peak in almost one day on December 20 (MJD 59203) followed by gradual spectral softening.
On December 23, the 2--10\,keV flux rapidly decreased to below the detection limit in one day \citep{Negoro2020} from $0.117\pm0.016$\,ct\,cm$^{-2}$\,s$^{-1}$ at MJD 59205.89 (the center time of the 6-hr bin) to $-0.003\pm 0.009$\,ct\,cm$^{-2}$\,s$^{-1}$ at 59206.86 (we note that background subtraction may result in nonphysical negative count rates).
However, we note that the light-curve obtained during the first \nustar\ observation performed on December 23 indicates a flux increase during the 4 hour observation (see \S \ref{sec:nustar}).
%%%% N.H. added on Oct. 20
This suggests that the rapid decrease observed by MAXI on December 23 is not simply due to the occultation by the companion. 
%%%%
This flux drop continued until around December 26.0 (MJD 59209.0), and the 4--10 keV flux increased again, possibly prior to an increase in the 2--4\,keV flux. 

We plot long-term variations in the flux of \maxi\ in Figure~\ref{fig:maxilclong}.
The data when the source is hidden by the Space-X Crew-1 spacecraft are not used and shown in grey.
GIS images for the duration show extended enhancement around the source region. 
The origin of the enhancement is unknown, but it is difficult to consider that it is intrinsic to the source because of the periodicity closely connected with the interference and the extended image even though there is still some ambiguity in the shadow shape of the spacecraft.

We also show the 2--10\,keV light-curve with a logarithmic time axis from the time December 22 04:48 (MJD59205.2 $\equiv T$) in the inset panel in Fig.~\ref{fig:maxilclong}.
We fit the decay with three different models: a linear decay, an exponential decay, and a power law decay. The linear fit to the putative rapid decay starting at $t_0 (\equiv t- T)=$ 0.69 d (=MJD 59205.89) is shown by the orange solid line. After $t_0 = 3.92$\,d the flux exhibited an exponential (blue curve) or a power law (green) decay.
Fitting the data for $t_0 \ge 3.92$\,d with an exponential function gives a time constant of $28.0\pm9.4$\,d (shown by the blue solid line and  its extrapolation by the dotted line). 
Fitting the flux decrease from $t_0 \ge 0.69$\,d to a power law function, we obtained a power law index of $-0.97\pm0.08$, shown by the green solid and dotted line, however we note that the result of the power law fit depends strongly on the choice of $T$.

\section{\nustar\ Observations}\label{sec:nustar}

\begin{deluxetable*}{ccccc}
\tablenum{1}
\tablecaption{Log of \nustar\ observations presented in this paper. \label{tab:obs_log}}
\tablewidth{0pt}
\tablehead{
\colhead{Observatory}   &   \colhead{Observation ID}    & \colhead{Start time (UTC)}    & \colhead{Stop time (UTC)} & \colhead{Exposure (s)}}
\startdata
\nustar                 &   90601340002 (NuObs~1)       & 2020-12-23 02:53:36           &  2020-12-23 07:06:26      & 9805                    \\
\nustar                 &   90601341002 (NuObs~2)       & 2020-12-31 09:08:34           &  2021-01-01 03:40:01      & 36395                    
\enddata
\end{deluxetable*}

\begin{figure*}[t!]
\begin{center}
\includegraphics[width=0.46\textwidth]{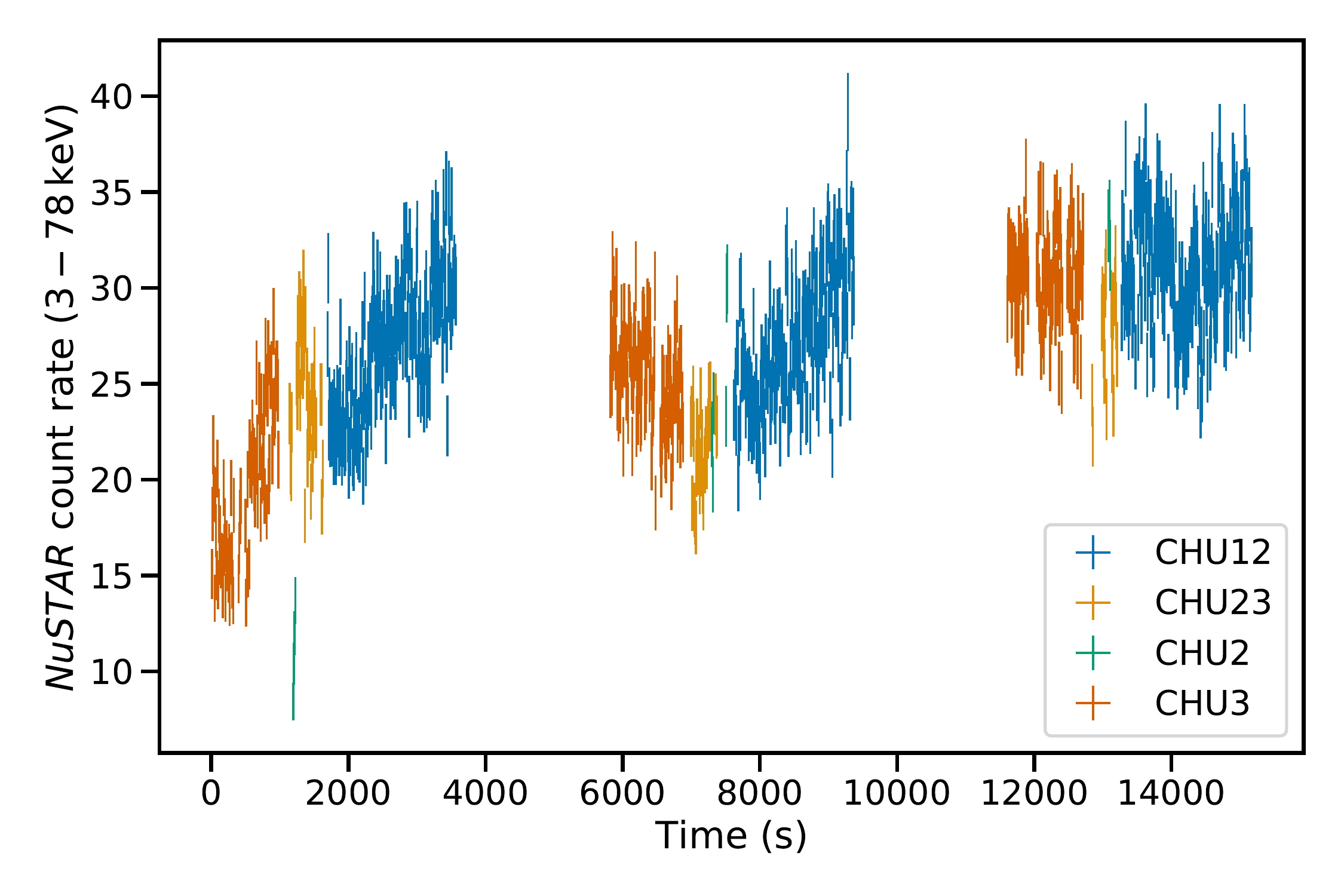}
\includegraphics[width=0.46\textwidth]{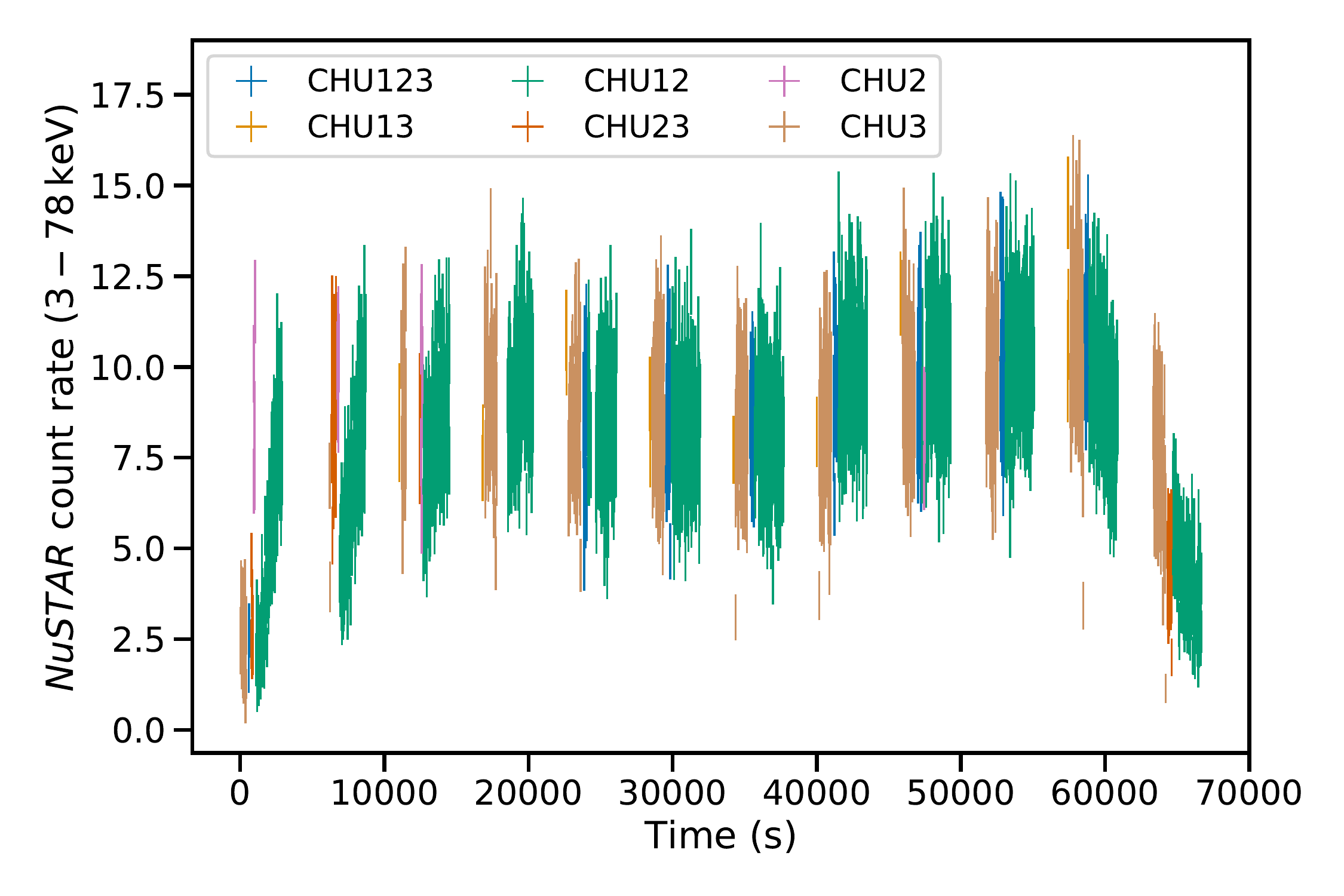}
\caption{Background-subtracted 3-78\,keV \nustar\ light-curve for each of the two observations, NuObs~1 (left) and NuObs~2 (right). The light-curves have been split into different CHU combinations. Time is measured in seconds since the observation start times listed in Table \ref{tab:obs_log}.
\label{fig:nustar_lc}}
\end{center}
\end{figure*}

\nustar, launched in June 2012, is the first high-energy focusing X-ray telescope \citep{Harrison2013}. It is composed of two focal planes, FPMA and FPMB, each paired with a set of focusing optics with a focal length of 10\,m. The focal planes are each made up of 4 pixelated Cadmium Zinc Telluride (CZT) detectors, bonded to a set of custom readout electronics. \nustar\ has a resulting bandpass of 3--78\,keV, making it a uniquely powerful tool for studying hard X-ray emission.

\maxi\ was first observed by \nustar\ on 2020 December 23. This observation was taken with the goal of localizing the source, and was therefore broken down into 5 pointings, forming a mosaic which tiled the error region reported by MAXI. The source was detected by \nustar\ in the first of these pointings (OBSID 90601340002, PI Fiona Harrison), which had an exposure time of 9.8\,ks. \nustar\ followed up on \maxi\ about a week later on 2020 December 31 (OBSID 90601341002, PI Fiona Harrison), after the source had dimmed. This observation had an exposure time of 36\,ks. Hereafter we refer to these observations as NuObs~1 and NuObs~2, respectively. Table \ref{tab:obs_log} lists the \nustar\ observations as well as their start and stop times and their total exposure times.

During both \nustar\ observations, \maxi\ was in close angular proximity to the Sun. This limits the aspect reconstruction as the primary Camera Head Unit (CHU) that \nustar\ uses to project counts onto the sky is blinded by the Sun. Instead, the ground software must make use of CHUs 1, 2, and 3 which are attached to the spacecraft bus. As a result, Mode 1 scientific data was unavailable. Instead, we analyzed Mode 6 scientific data, with which the source image cannot be perfectly reconstructed. This produces a source image with multiple centroids, each corresponding roughly to a different combination of CHUs. We reprocessed the unfiltered event files using NuSTARDAS v2.0.0 and CALDB v20200826. Next, we split the cleaned Mode 6 event files into event files corresponding to different combinations of CHUs 1, 2, and 3 using \code{nusplitsc} in the strict splitting mode. Figure \ref{fig:nustar_lc} shows the background-subtracted light-curve for each \nustar\ observation, split into different CHU combinations.

\begin{figure}
\begin{center}
\includegraphics[width=0.46\textwidth]{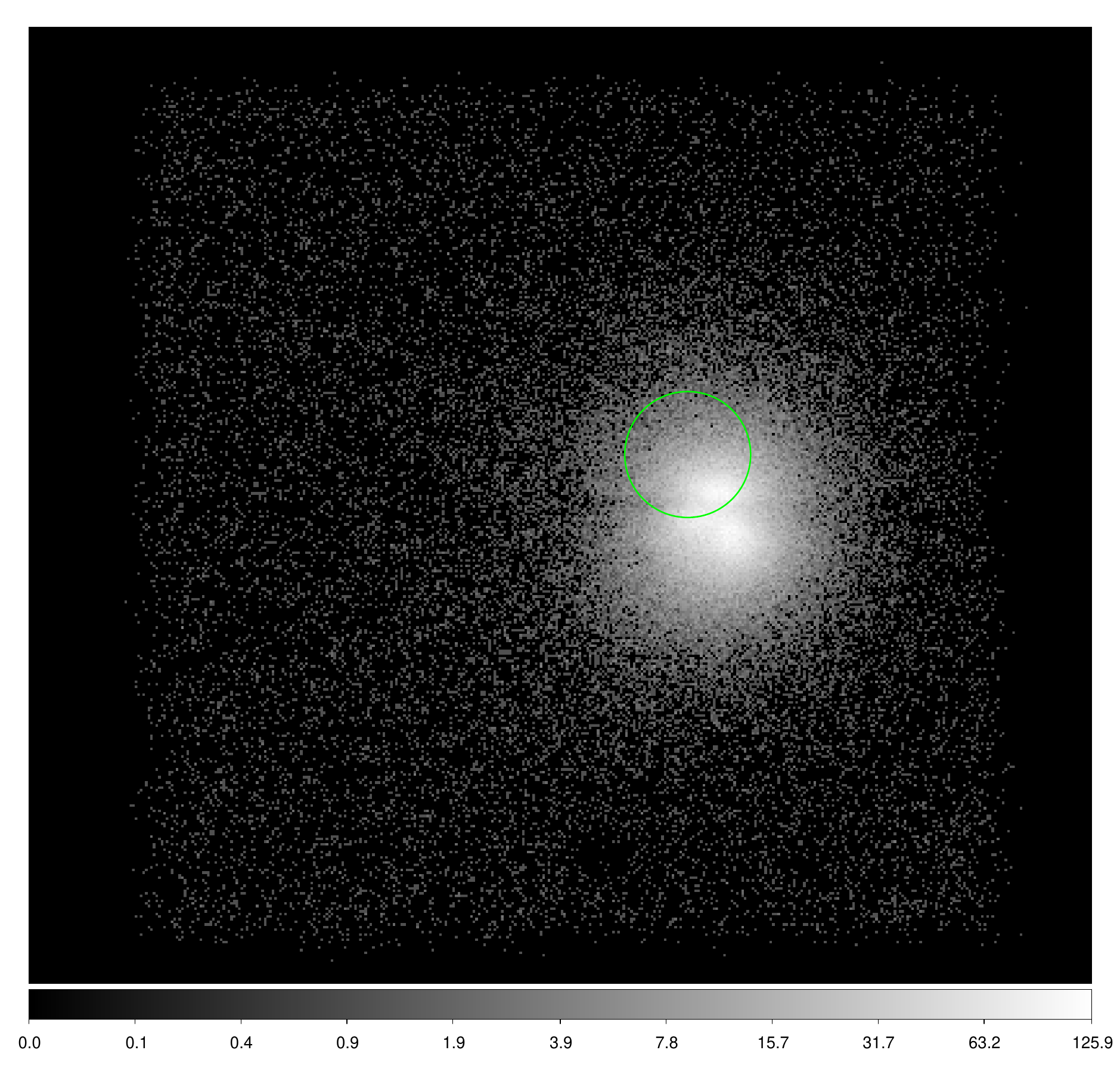}
\caption{Summed Mode 6 data (all CHU combinations) from NuObs~1. The color bar is shown in units of counts. The source landed on the gap between detectors, and due to poor image reconstruction, multiple peaks in the count distribution are apparent. The green circle, which has a radius of $1^{\prime}$, represents the ASCA error region for source AX J184848--0129 as reported by \citet{Sugizaki2001}.
\label{fig:ds9_view}}
\end{center}
\end{figure}

For the purpose of extracting spectra, we used DS9 \citep{Joye2003} to select circular source and background regions for each CHU combination. For NuObs~1, we chose source regions with radii equal to $60^{\prime\prime}$ and background regions with radii equal to $90^{\prime\prime}$. The centers of the source regions were determined using the automated centroid detection algorithm provided by DS9. During this observation, the source fell on the gap between detectors. Therefore, the source centroid fell on a different detector for different CHU combinations (see Figure \ref{fig:ds9_view}). We chose background regions which fell on the same detectors as the source centroid for each CHU combination. This was not a concern when choosing source and background regions for NuObs~2 because the source did not fall near the chip gap. Due to the lower count rate and better pointing during this observation, we chose source regions with radii of $45^{\prime\prime}$ in order to reduce the contribution of background counts, and we again chose background regions with radii of $90^{\prime\prime}$.

Using \code{nuproducts} we extracted scientific products for each of these CHU combinations, essentially treating each as a separate observation. We then used the routine \code{addarf} to sum the resulting Ancillary Response Files, and the routine \code{addspec} to sum the resulting source spectra, background spectra, and Response Matrix Files. The final result for each observation was a single source spectrum, background spectrum, RMF, and ARF for each focal plane module. We used the spectral fitting package \code{Xspec} \citep[v12.11.1][]{Arnaud1996} to analyze the spectra. In order to determine best fit models, we used the $W$ fit statistic \citep{Wachter1979}, and the full \nustar\ band, 3--78\,keV, was used for spectral fitting. Spectra were binned using the optimal binning procedure described by \cite{Kaastra2016}. Spectra shown in figures have been further rebinned for clarity such that each bin has a significance of at least $5\sigma$, with the exception of the highest-energy bins, each of which has a significance between $3\sigma$ and $4\sigma$. All errors quoted represent $90\%$ confidence intervals, and upper/lower limits represent $99\%$ confidence intervals, unless otherwise stated.

We made significant use of the Python modules Scipy \citep{Scipy2020}, Astropy \citep{Astropy2013,astropy2018}, Matplotlib \citep{Hunter:2007}, Corner \citep{corner}, and Stingray \citep{Huppenkothen2019}. We used Astropy to easily read and write NuSTAR event and light curve files, we used Scipy to perform various calculations and curve fitting, and we used Stingray to calculate and analyze power density spectra. We used Matplotlib and Corner to produce plots of spectra, light-curves, power density spectra, and parameter distributions.

\subsection{Spectral variability} \label{sec:spectra}

\begin{figure}[t!]
\begin{center}
\includegraphics[width=0.46\textwidth]{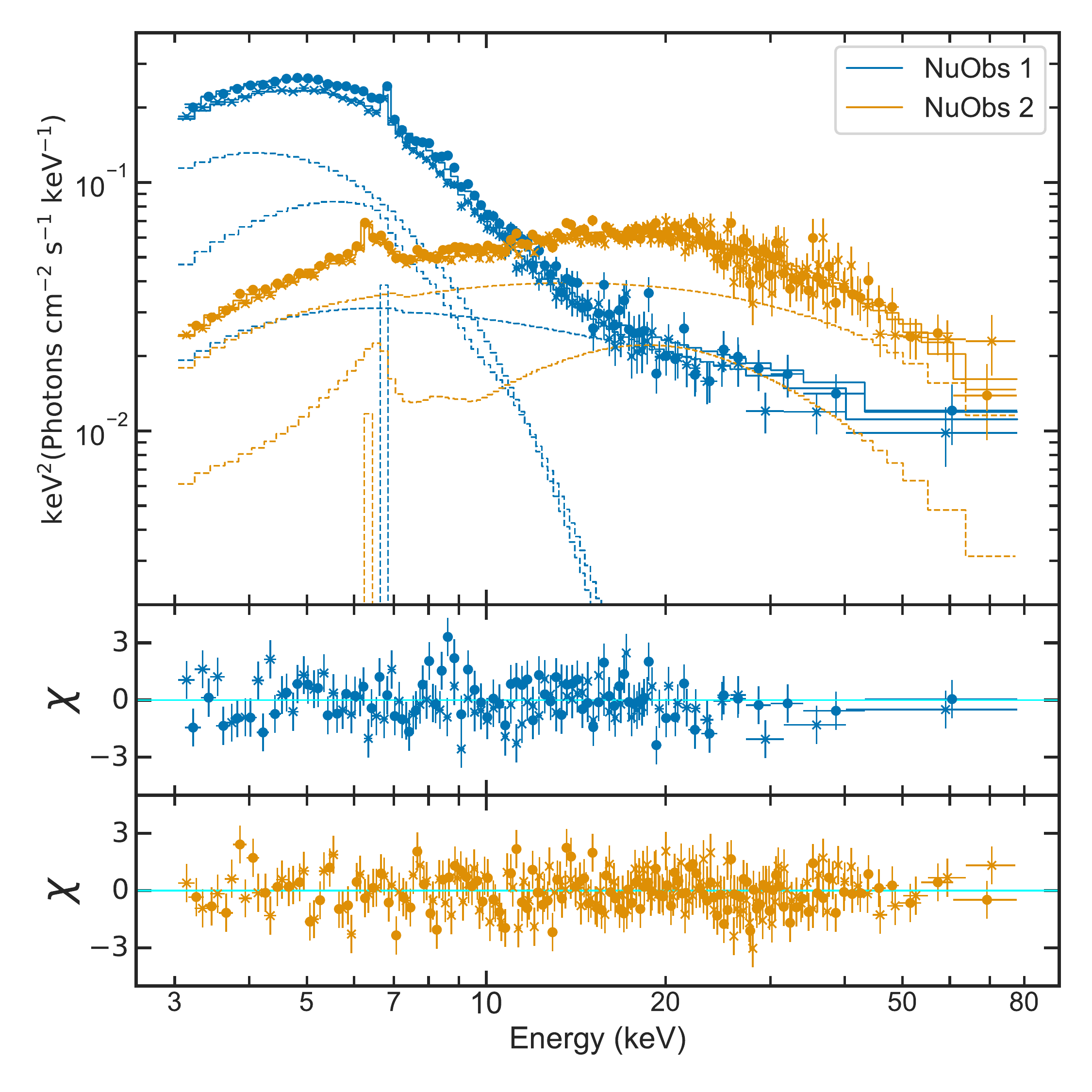}
\caption{Soft state (blue) and hard state (orange) spectra with best fit models and model components shown with solid and dashed lines, respectively. FPMA spectra are shown with x's while FPMB spectra are shown with filled circles. The soft state spectrum is described well by a Comptonized disk blackbody, while the hard state spectrum does not require a disk component. Instead, the hard state spectrum is described well by a power law with a high energy cutoff. Both models are improved by the addition of relativistic disk reflection as well as a narrow Fe line component. The lower two panels show the residuals resulting from the model fits in units of $\chi=\mathrm{(data-model)/error}$.
\label{fig:all_spectra}}
\end{center}
\end{figure}

\begin{figure}[t!]
\begin{center}
\includegraphics[width=0.45\textwidth]{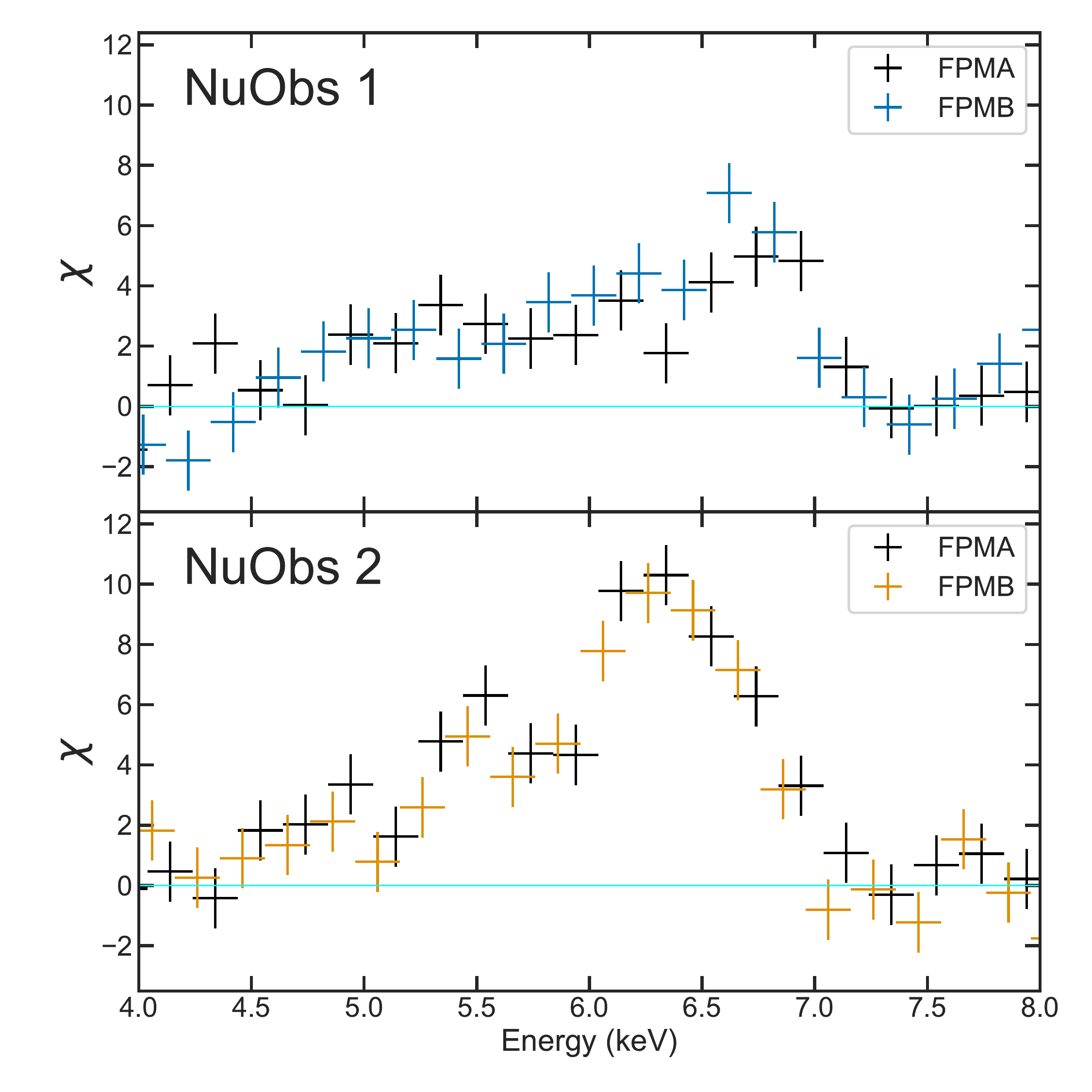}
\caption{Broad Fe emission profiles in the soft state (top) and hard state (bottom) spectra. These residuals were produced by fitting the spectra to a simplified continuum model while ignoring data points in the range $5-10$\,keV. Significant changes to the shape of the profile can be seen between the two states with the peak growing stronger and shifting to lower energies during the hard state.
\label{fig:Fe_lines}}
\end{center}
\end{figure}

\begin{deluxetable*}{cccc}
\tablenum{2}
\tablecaption{Spectral parameters determined by performing a joint fit of the soft and hard state spectra in Xspec. \label{tab:relxilllp_gauss}}
\tablewidth{0pt}
\tablehead{
\colhead{Model component}                       &   \colhead{Parameter}     & \colhead{Soft State (NuObs~1)} & \colhead{Hard State (NuObs~2)}}
\startdata
\noalign{\smallskip}
\code{tbabs}                                    & $N_{\rm H}$ ($10^{22}\,\mathrm{cm}^{-2}$)         & $4.2^{+0.2}_{-0.1}$       & $6.1 \pm 0.2$             \\
\noalign{\smallskip}
\hline
\multirow{3}{*}{\code{gauss}}                   & $\mu$ (keV) 				                        & $6.8 \pm 0.1$             & $6.3 \pm 0.1$             \\
                                                & $K$ ($10^{-5}\mathrm{Photon\,cm^{-2}\,s^{-1}}$)   & $17.7\pm0.4$ 	            & $6.26^{+0.05}_{-0.07}$    \\
                                                & Equivalent Width (eV) 			                & $42 \pm 3 $               & $42 \pm 1$                \\
\noalign{\smallskip}
\hline
\multirow{2}{*}{\code{diskbb}}                  & $kT_\mathrm{in}$ (keV)  	                        & $1.38^{+0.02}_{-0.03}$    & \nodata   	            \\
                                                & $(R_{\mathrm{km}}/D_{10})^{2}\cos{i}$             & $6.4^{+0.2}_{-0.3}$       & \nodata                   \\
\noalign{\smallskip}
\hline
\multirow{3}{*}{\code{nthcomp/cutoffpl}}        & $\Gamma$ 						                    & $2.42 \pm 0.04$           & $1.58^{+0.01}_{-0.02}$    \\
                                                & $E_\mathrm{cut}$ (keV)                            & \nodata                   & $28.9 \pm 0.4$            \\
                                                & Norm ($10^{-3}$)                                  & $9.6 \pm 0.4$             & $21.3^{+0.7}_{-0.8}$      \\
\noalign{\smallskip}
\hline
\multirow{8}{*}{\code{relxillNS/relxilllp}}     & $h$ ($R_\mathrm{g}$)	                            & \nodata                   & $5.7 \pm 0.2$             \\
                                                & $a$ 			                                    & \multicolumn{2}{c}{$0.967 \pm 0.013$}                 \\
                                                & Inclination (deg)                         & \multicolumn{2}{c}{$26.4 \pm 0.5$} 	                \\
                                                & $R_\mathrm{in}$ ($R_\mathrm{g}$)                  & $2.9 \pm 0.2$             & $7.90^{+0.15}_{-0.16}$    \\
                                                & $\log(\xi/\mathrm{erg\,cm\,s^{-1}})$		        & $2.58 \pm 0.06$	        & $2.77^{+0.04}_{-0.02}$    \\
                                                & $A_{\mathrm{Fe}}$ (Solar)                         & \multicolumn{2}{c}{$0.77^{+0.02}_{-0.01}$}            \\
                                                & $\log(N)$ ($\mathrm{cm^{-3}}$)                    & $<15.3$                   & \nodata                   \\
                                                & $f_\mathrm{refl}$\tablenotemark{\small a}         & $0.61 \pm 0.01$           & $0.83^{+0.02}_{-0.03}$    \\
\noalign{\smallskip}
\hline
\hline
\noalign{\smallskip}
\multicolumn{2}{c}{$F_\mathrm{bol}(10^{-10}\,\mathrm{erg\,cm^{-2}\,s^{-1}})$\tablenotemark{\small b}}   & $6.9\pm0.1$           & $2.85\pm0.04$             \\
\multicolumn{2}{c}{$L_\mathrm{bol}(10^{35}\,\mathrm{erg\,s^{-1}})$\tablenotemark{\small c}}             & $9.1\pm0.1$           & $3.73\pm0.05$             \\
\multicolumn{2}{c}{$W/\mathrm{d.o.f.}$}                                                             & \multicolumn{2}{c}{834/695}                           \\
\noalign{\smallskip}
\enddata
\tablenotetext{\small a}{\footnotesize We determined the reflection fraction by first fitting with the direct emission components included in the \code{relxill} models (i.e. with a positive value of $f_\mathrm{refl}$), then we determined the normalizations of the direct components by freezing the reflection fractions and separating the direct and reflected components (with a negative value of $f_\mathrm{refl}$).}
\tablenotetext{\small b}{\footnotesize Unabsorbed bolometric ($0.1-100\,\mathrm{keV}$) flux.}
\tablenotetext{\small c}{\footnotesize Bolometric ($0.1-100\,\mathrm{keV}$) luminosity assuming a distance of $3.3\,\mathrm{kpc}$.}
\end{deluxetable*}

\nustar\ revealed a stark spectral change between the first and second observations. The spectra are shown in Figure \ref{fig:all_spectra} along with their best-fit models. We found that the spectrum during NuObs~1 was significantly softer than that observed during NuObs~2. We therefore refer to the former as the ``soft state" and the latter as the ``hard state". The spectra are described remarkably well by models which are frequently used to model the soft and hard states of accreting black holes and neutron stars: the continuum emission during the soft state is described well by an absorbed disk blackbody (\code{diskbb}) and a power law with index $\Gamma$, while the continuum emission during the hard state is described well by an absorbed power law with a high-energy cutoff, $E_\mathrm{cut}$, which we modeled using \code{cutoffpl}. We modeled absorption using the \code{tbabs} model. We used cross-sections provided by \citet{Verner1996} and abundances provided by \citet{Wilms2000}.

Each of the spectra exhibits broad emission lines around 6.4\,keV corresponding to Fe K$\alpha$. Figure \ref{fig:Fe_lines} shows the Fe line profiles which result from fitting the continuum emission while ignoring data bins between 5 and 8\,keV. The broad, asymmetrical structure of the Fe line profiles led us to add relativistic disk reflection to the continuum models described above. We added a reflected component using \code{relxilllp} \citep{Dauser2014,Garcia2014} in order to self-consistently model emission originating from a hot corona with a lamppost geometry which irradiates and is reprocessed by a thin disk. We tied the power law index and high energy cutoff\footnote{Because the soft state does not require a high-energy cutoff, we fixed the cutoff energy at 1000\,keV, far beyond the upper edge of the \nustar\ bandpass.} of each power law component to those of \code{relxilllp}. We found that this model was able to describe the the spectrum well, achieving a fit statistic of $W/\mathrm{d.o.f.}=449/323$ and $W/\mathrm{d.o.f.}=420/373$ (and a reduced Chi-squared test statistic of $\chi^2_{\nu}=0.98$ and $\chi^2_{\nu}=1.01$) for the soft and hard states, respectively. However, the Fe line emission could not be fully accounted for, and residuals indicated the presence of a narrow line component. We therefore added a Gaussian component near 6.4\,keV, and we froze its width at $\sigma=10^{-5}$\,keV, much narrower than the instrumental energy resolution. We allowed the strength and the centroid of the line to vary while fitting, and we found that the addition of this narrow line improved the fit by $\Delta W=-17.5$ and $\Delta W=-10.6$ for the soft and hard states, respectively, while decreasing the number of degrees of freedom by 2. We also found that the addition of this component led to better constraints on other parameters, such as the spin and the inclination, while still remaining consistent with their previous estimates. 

In order to determine the statistical significance of the narrow components, we simulated 5000 spectra each for the soft and hard states, originating from the reflection models without narrow components. We then added a narrow component, again leaving the centroid energy and line strength free and fixing the width at $\sigma=10^{-5}$\,keV. By calculating the resulting improvement in the fit statistic for each simulation, we arrived at a false positive rate for the addition of a narrow component. We found that for the soft state, only one of these simulations was improved by more than $\Delta W=-17.5$, and for the hard state, 23 out of the 5000 simulations were improved by more than $\Delta W=-10.6$. In other words, we estimate a statistical significance of $3.7\sigma$ and $2.8\sigma$ for the addition of a narrow Fe component in the soft and hard state observations, respectively.

Finally, prompted by the clear presence of a hot thermal component, when modeling the soft state spectrum we replaced \code{relxilllp} with the sum of \code{nthcomp} \citep{Zdziarski1996,Zycki1999}, which models Comptonization of seed photons originating from a blackbody,  and \code{relxillNS} \citep{Garcia2021}, which models reflection of a blackbody spectrum rather than a power law. We tied the disk blackbody temperature to that of the seed photon temperature for both \code{nthcomp} and \code{relxillNS}, and we fixed the electron temperature of \code{nthcomp}, $kT_e$, at $1000$\,keV. We found that the fit was not affected by the type of seed blackbody specified for \code{nthcomp}, so for consistency we used a disk blackbody seed spectrum. We found that this model improved the fit ($\Delta W=-10$, $\Delta\mathrm{d.o.f.}=1$) when compared to the \code{relxilllp} model. We also tried modeling the hard state spectrum using a combination of a Comptonized disk blackbody with a reflected Fe line (\code{diskbb $+$ nthcomp $+$ rellinelp}). We found that this model struggled to describe the hard state spectrum ($W/\mathrm{d.o.f.}=472/372$), particularly at high photon energies ($>20$\,keV). It resulted in a photon index of $\Gamma=1.8$ and an electron temperature of $kT_e=9.1$\,keV.

Having arrived at a suitable model for the spectra during both the soft (\code{diskbb $+$ nthcomp $+$ relxillNS $+$ gaussian}) and hard (\code{cutoffpl $+$ relxilllp $+$ gaussian}) states, we performed a Markov Chain Monte Carlo analysis in \code{Xspec} in order to explore the parameter space in detail. We used the Goodman-Weare algorithm \citep{Goodman2010} with a chain length of $10^6$ and a burn-in length of $10^5$. We initialized all walkers in a Gaussian distribution about the best fit parameters, with $\sigma$ defined by the covariance matrix resulting from least-squares fitting. All spectral parameters are therefore given as the median values of the final posterior distributions, and the errors represent the bounds between which $90\%$ of the samples lie.\footnote{See \citet{Hogg2018} for a discussion of best practices when interpreting the results of MCMC analyses.} We found that both models recovered a high spin, $a>0.7$, and an inner disk inclination angle around $25$\,deg. However there was significant degeneracy such that each model had valid solutions with different values for spin, disk inclination, inner disk radius, and other key parameters (see Appendix \ref{sec:individualfits} for details regarding the individual fits). In order to break this degeneracy, we performed a joint fit wherein the inner disk inclination, the spin parameter, and the iron abundance were tied between the two spectral states.

Table \ref{tab:relxilllp_gauss} lists the median parameters for the joint fit resulting from the MCMC analysis described above, as well as the $W$-statistic of the best fit. Figure \ref{fig:all_spectra} shows the spectra as well as the best fit models and residuals for both the soft and the hard states. By performing a joint fit, we successfully narrowed the parameter space, resulting in a consistent solution across both the soft and hard states with a spin of $a=0.967\pm0.013$ and an inclination of $i=26.4\pm0.5$\,deg. 

The Fe K$\alpha$ emission line profile (Figure \ref{fig:Fe_lines}) shows clear evolution of the region responsible for this emission, with the peak shifting towards lower energies by about $0.5$\,keV and the overall profile becoming more pronounced as the source evolved from the soft to the hard state. Interestingly, we found that the centroid of the narrow Fe line during the hard state was less than $6.4$\,keV at the $95\%$ confidence level. In other words, we detect a redshift in the narrow component at a significance of about $2\sigma$. We also fit the data while freezing the centroid at $6.4$\,keV given that this is still consistent with our measurement of $\mu=6.3\pm0.1$\,keV, but this resulted in a somewhat degraded fit ($\Delta W=3$; $\Delta \mathrm{d.o.f.}=1$), and it led to significantly looser constraints on parameters such as spin, inclination, and lamppost height. 

The evolution of the accretion disk is further reflected in the best-fit relativistic reflection models, which suggest a significant increase in the inner radius of the accretion disk. This could constitute evidence for disk truncation in the hard state. In order to determine whether the choice of spectral model affects the apparent evolution of the inner disk radius, we performed a joint fit wherein the disk reflection features of both the soft and hard state spectra were modeled using \code{relxilllp} (rather than using \code{relxillNS/relxilllp} for the soft/hard states, respectively). For this fit, we tied the spin parameter, the inner disk inclination, and the Fe abundance between the soft and the hard states. This resulted in a slightly larger value for the inner disk radius in both states -- $R_\mathrm{in}=3.4 \pm 0.2\,R_\mathrm{g}$ (where $R_\mathrm{g}=GM/c^2$) for the soft state and $R_\mathrm{in}=8.6^{+0.5}_{-1.1}\,R_\mathrm{g}$ for the hard state -- but the radii remained significantly different, indicating that the evolution of the inner disk radius which we observe is independent of the choice of \code{relxillNS} or \code{relxilllp}.

The narrow Fe line components may be interpreted as distant disk reflection. Therefore we investigated whether non-relativistic reflection modeling could describe this component. We chose to use another member of the \code{relxill} suite, \code{xillver} \citep{Garcia2010,Garcia2013} for this component. In this model, the \code{relxill} and \code{xillver} components can be interpreted as reflection from the inner and outer regions of the disk, respectively. In the soft state, distant reflection, using either \code{xillver} or \code{xillverNS} to model reflection of the power law component or the blackbody component, respectively, is not able to fit the data as well as a simple Gaussian and results in a change in the fit statistic of $\Delta W=10$. 

For the hard state, replacing the Gaussian emission line with the distant reflection model did not change the fit statistic appreciably, but we found that the resulting parameters show significant degeneracies, such as between the disk inclination, the power law photon index, and the lamppost height. We tied the photon index, $\Gamma$, and the cutoff energy, $E_\mathrm{cut}$, of the distantly reflected power law to those of the power law reflected from the inner disk, and we tied the iron abundance of the inner and outer disk but we allowed the ionization of the two components to vary independently. Given that we measured a centroid of $\mu = 6.3\pm0.1\,\mathrm{keV}$, slightly lower than the rest frame energy of the neutral Fe K$\alpha$ line, when modeling the narrow emission component using a Gaussian component, we also allowed the redshift of the outer disk component represented by \code{xillver} to vary.

We found that this model was not sensitive to the inclination of the outer disk. When allowed to vary independently of the inner disk inclination the model preferred a high outer disk inclination of $i_\mathrm{out}=79^{+6}_{-11}$\,deg ($\Delta i \equiv i_\mathrm{out} - i_\mathrm{in} = 57\pm10$\,deg), but tying the inner and outer inclinations did not affect the fit statistic and resulted in an inclination which was closer to that resulting from the joint fits, $i=22^{+4}_{-5}$\,deg. In addition to a slightly smaller inclination angle, adding the \code{xillver} component while tying the inner and outer disk inclinations also resulted in an increased inner disk radius ($R_\mathrm{in}=14^{+6}_{-5}\,R_\mathrm{g}$). 

We also found that allowing the redshift of the \code{xillver} component to vary improved the fit by $\Delta W=-4$ compared to the same model with the redshift frozen at a value of zero. We measure a redshift of $z<3.4\times10^{-2}$ at the $99\%$ confidence level, corresponding to an upper limit on the line-of-sight velocity of the emitting region of $v<10000\,\mathrm{km\,s^{-1}}$, assuming a rest frame energy of $6.4$\,keV. We found that performing the same calculations using the posterior distribution of Gaussian centroids resulting from the joint fit produces upper and lower limits differing from these by less than $10\%$. The fit is slightly degraded by fixing the redshift to zero, but the main effect on the parameters is a decrease in the inner disk inclination by a few degrees. We note that the source is unlikely to experience a strong redshift due to the motion of the binary system, and it certainly does not experience cosmological redshift, but rather any observed strong redshift is more likely to be caused by motion of the emitting region itself. In Section \ref{sec:narrow_lines}, we discuss one scenario -- that of ionized outflows -- which could result in red- and blue-shifting of narrow emission.

\subsection{Timing analysis} \label{sec:nustar_timing}

\begin{figure}
\begin{center}
\includegraphics[width=0.46\textwidth]{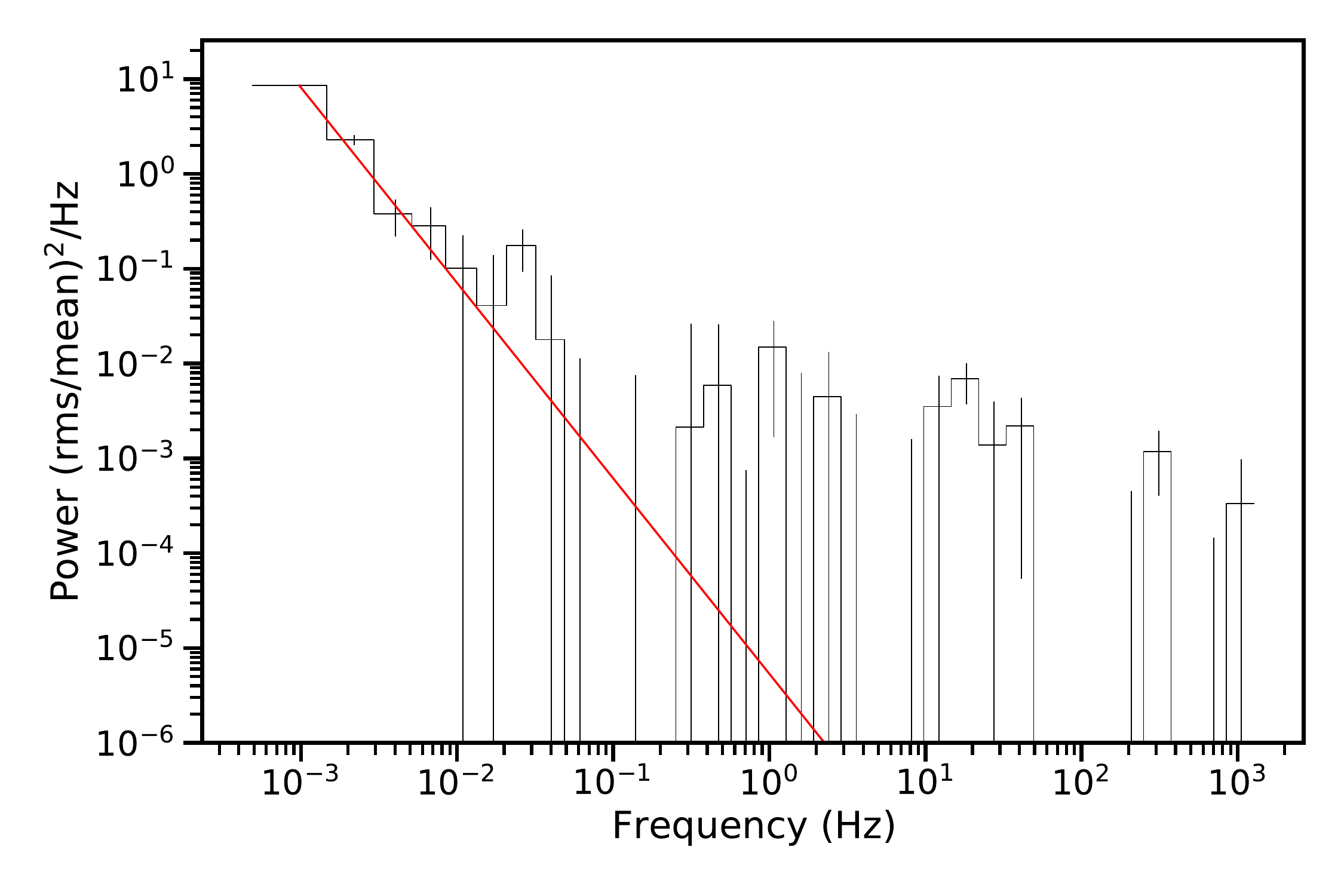}
\caption{Averaged cospectrum calculated for the hard state \nustar\ observation, NuObs~2, using only CHU12 data. The cospectrum has been rebinned for clarity. The best-fit power law is shown in red. We did not find evidence for QPOs or pulsations.
\label{fig:nustar_cospectra}}
\end{center}
\end{figure}

\begin{deluxetable*}{cccc}
\tablenum{3}
\tablecaption{Timing properties  of the source during each \nustar\ observation. \label{tab:timing_tab}}
\tablewidth{0pt}
\tablehead{
\colhead{Observation} &  \colhead{rms ($10^{-3}$--$10^{-1}$\,Hz)}    & \colhead{rms ($0.1$--$10$\,Hz)}   & \colhead{$\gamma$}}
\startdata
NuObs~1     & $<8\%$                &  $<15\%$      & \nodata                    \\
NuObs~2     & $12\pm2 \%$           &  $<14\%$      & $-2.1\pm0.1$                    
\enddata
\end{deluxetable*}

We investigated the source variability by producing cospectra \citep{Bachetti2015} for each of the \nustar\ observations. We first corrected the photon times of arrival for the motion of \nustar\ using \code{barycorr}. We specified the {\em Chandra} source position reported by \citep{2021ATel14424....1C}, and we used the JPL planetary ephemeris DE-430. In order to minimize the impact of the motion of the source on the detectors on our timing analysis, we analyzed each of the CHU combinations separately. We split the event files into segments of length $1024$\,s, then binned the events with a time resolution of $2048^{-1}$\,s in order to probe variability between $10^{-3}$\,Hz and $10^3$\,Hz\footnote{Recent improvements to the calibration of \nustar\ have greatly improved its timing capabilities, allowing for precision down to the $\sim65\,\mathrm{\mu s}$ level \citep{Bachetti2021}}. This range of frequencies allowed us to investigate both the low-frequency noise, and to search for any hint of pulsations potentially arising from a neutron star source. For each segment, we calculated the cospectrum between FPMA and FPMB. We found that for both observations, only CHU combination CHU12 contained continuous good time intervals of length greater than 1024\,s, and therefore this was the only CHU combination for which we calculated cospectra. For each observation we then averaged all of the cospectra. Finally, we applied deadtime and background corrections by multiplying the averaged fractional rms normalized power by $(1-\tau(\overline{S+B}))^{-2}(\overline{S+B}/\overline{S})^2$, where $\tau=25\,\mathrm{ms}$ is the deadtime per event, $S$ and $B$ are the source and background rates, respectively, calculated from the event list, and bars indicate the geometric mean of the FPMA and FPMB count rates.

Table \ref{tab:timing_tab} shows the timing properties of the source during both observations. Because the source landed on the gap between detectors during NuObs~1, which is sure to exacerbate any systematic variability due to motion of the source on the focal plane, we do not present the cospectrum here. The quality of the data was not sufficient to perform modeling of the cospectrum, but we did not observe a clear flattening in the low frequency noise of the source. We were able to place an upper limit on the low-frequency noise of $\mathrm{rms}<8\%$ by integrating the measured power in the range $10^{-3}-10^{-1}$\,Hz. In order to estimate the continuum rms variability, we produced an averaged cospectrum with segment length $64$\,s (due to the shorter segment length, we were able to utilize the data from all CHU combinations except for CHU2) and sampled the resulting error distributions for the power in the range $0.1-10$\,Hz. We produced $10^5$ sample cospectra in this range, and calculated the total rms for each. The result was an upper limit, defined as the 99th percentile of this sample, of $\mathrm{rms}<15\%$.

The source did not fall on the chip gap during NuObs~2, and we therefore present the averaged, logarithmically rebinned cospectrum in Figure \ref{fig:nustar_cospectra}. Although there is clear red noise, similar to the previous observation we do not observe a low frequency turnover. We fitted the low-frequency ($<0.1$\,Hz) noise to a power law and found an index of $\gamma=-2.1 \pm 0.1$. This is consistent with the tail of a zero-centered Lorentzian, which is often used to describe the low-frequency noise in accreting black holes and neutron stars \citep{Belloni2002}. We integrated the low-frequency noise and obtained $\mathrm{rms}=12\pm2\%$, and we obtained an upper limit on the power integrated between $0.1-10$\,Hz of $\mathrm{rms}<14\%$ using the same method as described for the previous observation (in this case only CHU13 was excluded). 

Noting a potential feature in the cospectrum at $\sim2\times10^{-2}$\,Hz, we performed a search for quasiperiodic oscillations (QPO) by stepping logarithmically from $10^{-3}$\,Hz to $1024$\,Hz and fitting the unbinned hard state cospectrum to the power law described above plus a Lorentzian centered at each frequency step. For each frequency, we recorded the change in the $\chi^2$ fit statistic compared to the power law alone. Using this method, we did not find any significant QPO candidates. To determine whether the potential feature could be due to coherent pulsations, we also folded the events for each observation, summing all CHU combinations, at 10000 pulse frequencies between $2\times10^{-2}$\,Hz and $3\times10^{-2}$\,Hz and calculating the $Z^{2}_{1}$ statistic \citep{Buccheri1983} for each candidate frequency. For the soft state observation, NuObs~1, we found a peak in the $Z^{2}_{1}$ distribution at $\sim44$\,s with a significance of $2.1\,\sigma$, and for the hard state observation, NuObs~2, the highest peak was at $\sim42$\,s and had a significance of only $0.7\,\sigma$. We further searched for pulsations in each observation by producing cospectra for the frequency range $0.008$--$2048$\,Hz, again using the sum of all CHU combinations. We did not observe any significant peaks during either observation.

\section{Discussion} \label{sec:discussion}
\subsection{The nature of the accretor} \label{sec:NS_or_BH}
While there is no ``smoking gun" evidence for one particular type of compact object (for example, the detection of pulsations would prove the presence of a neutron star), the evidence we have presented thus far favors the case of a black hole X-ray binary. 

The soft state spectrum is described well by a Comptonized multi-temperature disk blackbody with relativistic reflection features, which may represent returning radiation \citep[c.f.][]{Connors2020,Connors2021,Lazar2021}, and the hard state spectrum is described by a reflected power law alone. Joint modeling using \code{relxilllp} and \code{relxillNS} indicates that the accreting object has a high spin, $a\approx0.97$. Given the relation $a=cJ/GM^{2}$, where $J$ is the angular momentum of the spinning compact object, and $M$ is its mass, and assuming a moment of inertia $I=\frac{2}{5}MR^{2}$ for a neutron star, we may estimate the maximum spin that a neutron star can attain. Assuming a canonical neutron star with $M=1.4M_{\odot}$ and $R=10$\,km, we arrive at a spin of $a\approx0.4\frac{\nu_\mathrm{spin}}{\mathrm{kHz}}$, meaning that even the most rapidly rotating neutron stars will not achieve spin parameters approaching that which we have measured. We consider this a compelling piece of evidence which favors the classification of \maxi\ as an accreting black hole -- explaining the broad Fe line in the case of an accreting neutron star would prove difficult. 

Additionally, using archival Chandra observations of the GC01 cluster, \citet{Hare2021} were able to place an upper limit of $\sim3.3\times10^{30}\,\mathrm{erg\,s^{-1}}$ on the quiescent luminosity in the 0.5--10\,keV band. Whereas accreting neutron stars tend to have quiescent luminosities around $10^{33}\,\mathrm{erg\,s^{-1}}$ \citep{Tsygankov2017}, the luminosity of \maxi\ prior to outburst is more in line with those measured for black holes, perhaps due to the presence of an event horizon rather than a boundary layer \citep{Garcia2001}.

The initial outburst observed by MAXI was very short-lived compared to many black hole X-ray binaries, which tend to show exponential decays with e-folding times of tens of days \citep{McClintock2006}, however similarly rapid outbursts are not unheard of \citep[c.f.][]{Maitra2006}. Of course, the count rate in the limited energy band does not always reflect the mass accretion rate, especially in cases where the source spectrum changes significantly. It is, however, interesting to note that some objects, e.g. gamma-ray bursts, exhibit power law decays similar to the flux decrease observed for \maxi\ by \nustar, MAXI, and Swift \citep{Miller2021}.

\subsection{Disk accretion at low apparent luminosity} \label{low_luminosity}
\maxi\ reached an unabsorbed bolometric luminosity of $\sim10^{36}\,\mathrm{erg\,s^{-1}}$ during the soft state, and dropped to a luminosity of $\sim4\times10^{35}\,\mathrm{erg\,s^{-1}}$ in the hard state. The general shape of the spectrum observed during each of these states is consistent with accretion onto a neutron star or a black hole, and despite the low luminosities, observations of broad emission features around $6.4$\,keV indicate the presence of an optically thick accretion disk which extends close to the central accretor and reaches a high temperature of $\sim1.4$\,keV. Models of accretion onto black holes predict the formation of an advection-dominated accretion flow, or ADAF, at mass accretion rates below a few percent of the Eddington limit, resulting in truncation of the disk at large radii \citep{Esin1997}. Indeed, observations have demonstrated that X-ray binaries tend to undergo state transitions at $\sim2\%$ of the Eddington limit \citep{Maccarone2003,Motlagh2019}. Given the inconsistency we observe between the low source luminosity and evidence for a small disk radius, we consider two effects which could suppress the observed luminosity: disk inclination and obscuration.

Reflection features -- the broad Fe line profile and excess ``Compton hump" above $10$\,keV -- allowed us to constrain the inclination of the inner regions of the disk as well as the inner disk radius. Reflection modeling of the soft state spectrum resulted in an inner disk radius of $R_\mathrm{in}=2.9 \pm 0.2R_\mathrm{g}$. Even for a neutron star accretor with mass $M=1.4M_\odot$, this implies an inner disk radius of about 6\,km, whereas the flux of the disk blackbody implies an effective inner disk radius of $R_\mathrm{in} \sqrt{\cos i}=0.8\,\mathrm{km}$, assuming a distance to the source of $3.3$\,kpc. Several effects, including gravitational redshift, Compton scattering of disk photons \citep{Kubota2004}, spectral hardening \citep{Shimura1995}, and boundary condition corrections \citep{Kubota1998}, will increase or decrease the actual value of the inner disk radius compared to the apparent radius, but in total these effects are unlikely to introduce a factor of more than order unity. 

Thus we are left with a contradiction between inner disk radius measurements which could presumably be resolved by assuming a high inclination, $i$, of the inner disk, resulting in a smaller projected disk flux from the point of view of an observer. While reflection modeling appears to rule this out, preferring an inclination between 20 and 30 degrees, it has been shown that the inclination inferred from reflection modeling can conflict drastically from the actual inclination of a system, possibly due to obscuration of the blue wing of the broad Fe line \citep{Connors2019}.

High inclination of the inner disk alone may not be able to explain the low flux of \maxi. We therefore also consider obscuration of the inner regions of the disk by the outer disk. In this scenario, the outer regions of the disk intercept the line of sight between the observer and the inner disk. If the outer regions of the disk have a clumpy structure, this will result in an effective partial coverer, and may be consistent with the change in column density we have measured between the soft and hard states. Alternatively, we may be observing emission from the inner disk which has been scattered off the outer regions of the disk on the side opposite the observer. 

A combination of these two effects has been proposed for the high mass X-ray binary, V4641 Sgr \citep{Koljonen2020}, which has similarly been shown to exhibit evidence for a disk extending close to the central black hole despite a low Eddington ratio \citep{Miller2002}. In fact, the hard state of \maxi\ shares a number of similar characteristics with V4641 Sgr. Both exhibit power law spectra with indices between 1 and 2 and relatively low cutoff energies ($<30$\,keV). Additionally, both systems undergo short outbursts with durations of tens of days or less \citep{Wijnands2000,Revnivtsev2002}, and on shorter timescales they both exhibit relatively featureless power spectra with red noise components described by a power law of slope $\sim-2$ \citep{Maitra2006}. However, the amplitude of variability in the 0.1--10\,Hz range that we have observed, while poorly constrained, is somewhat lower than that of V4641 Sgr \citep{Maitra2006}. In both of these systems, reprocessing of emission from the central regions of the accretor could help to explain this behavior by ``smearing out" variability on short timescales. 

Given the low luminosity as well as the very short outburst duration, we also suggest the possibility that the source may reside in a compact binary. A smaller-than-usual disk could lead to short outbursts, and depending upon the orbital period, the mass accretion rate may be low \citep{Deloye2003}. It is interesting to note that black hole binaries undergoing mass transfer in globular clusters are likely to harbor both low-mass donors and relatively low-mass black holes \citep{Kremer2018}. Because we are unable to detect the orbital period, and due to the fact that a compact binary would imply a small companion, which is contradicted by the potential NIR counterpart of the system, we consider this case speculative. However, given the rarity of known black holes in compact binaries, it is an interesting possibility nonetheless and one which warrants further observations of the source.

\subsection{Evolving narrow line emission} \label{sec:narrow_lines}
We found that the narrow Fe lines observed in the soft and hard state spectra have different centroid energies, evolving from $\sim6.8$\,keV in the soft state to $\sim6.3$\,keV in the hard state. This change likely corresponds to a change in the ionization of the emitting region, perhaps originating from the outer regions of the accretion disk, with neutral or lowly-ionized Fe\,{\scriptsize I}--{\scriptsize XVII} responsible for the line in the hard state and highly ionized Fe\,{\scriptsize XXV}--{\scriptsize XXVI} responsible for the line in the soft state \citep{Garcia2013}. Indeed, the significantly higher thermal X-ray flux in the Fe K$\alpha$ complex energy range during the soft state would lead to a higher ionization.

Rather than an evolution of the emitting region, it is also possible that narrow emission from neutral Fe was actually present during both states, but that it did not increase drastically in flux during the soft state and was therefore not detectable above the thermal-dominated continuum, which was significantly higher in flux around $6.4$\,keV as compared with the hard state. Indeed, when \maxi\ was observed by NICER during a second, less pronounced increase in flux, \citet{Miller2021} not only reported that the spectrum was described well by a blackbody with temperature $kT=1.0$\,keV, indicating that the source had again entered the soft state, but the authors also reported both the detection of a prominent, narrow emission line at $6.7$\,keV with flux $K=1.5\times10^{-5}\,\mathrm{photon\,cm^{-2}\,s^{-1}}$, as well as a tentative detection of a weaker line at $6.4$\,keV. 

In order to test whether the lower-energy line could have been present but undetected during the \nustar\ observation of the soft state, we added a second Gaussian component to the soft state in our joint spectral model. We froze the centroid at $\mu=6.4$\,keV and the width at $\sigma=10^{-5}$\,keV. We allowed the flux of this component to vary, then we calculated the $99\%$ confidence upper limit as defined by a change in fit statistic of $\Delta W=6.63$. The resulting upper limit was $K<7.7\times10^{-5}\,\mathrm{photon\,cm^{-2}\,s^{-1}}$. In other words, we cannot rule out the possibility that neutral Fe emission was present in the soft state at a similar flux level to that observed in the hard state.

One alternative scenario to that of differing ionization is an ionized outflow. Given that we do not observe clear absorption features which might indicate disk winds, such an outflow may instead take the form of a jet launched during the short outburst observed by MAXI. In this case, the narrow emission lines may originate from fast-moving, ionized blobs of plasma on opposite sides of the accretor along the jet axis. One would then expect to see a pair of components, one of which represents the blue-shifted jet component moving towards the observer, and the second of which originates from the red-shifted jet component moving away from the observer. Due to relativistic beaming, the blue-shifted component would have a higher observed flux than that of the red-shifted component. Additionally, depending on the orientation of the system one may expect a delay between the appearance of the blue component and that of the red component due to the difference in light-travel time from each of the emitting blobs. \citet{Migliari2002} observed a pair of Fe lines during Chandra observations of SS~433, which they were able to spatially associate with extended jet emission. They determined that the Fe line emitting regions were located at a distance of $>10^{17}$\,cm from the central accretor. For a source with low or moderate inclination, this would result in a light-travel time between the two lobes of $30$--$40$\,days.

Indeed, we have shown that the $6.8$\,keV line had a higher flux than, and was observed prior to, the $6.3$\,keV line. If the two lines originated from regions with similar ionization states, and the difference in energy is solely due to blue- and red-shifting, then we may calculate the velocity of the narrow line emitting regions:

\begin{equation}
    \frac{E_\mathrm{obs}}{E_\mathrm{rest}} = \frac{1}{\gamma\left(1-\beta\cos{\theta}\right)}
\end{equation}

\noindent where $E_\mathrm{obs}$ is the energy of the line in the frame of the observer, $E_\mathrm{rest}$ is the energy of the line in the rest frame of the emitting region, $\beta$ is the ratio of the velocity of the emitting region to the speed of light, $\gamma$ is the Lorentz factor, and $\theta$ is the angle at which the emitter is moving with respect to the line of sight. If we assume that the velocities of the red and blue components are equal in magnitude, the result is $\beta\cos{\theta}\approx0.038$. Given that we have measured a moderate inclination via spectral analysis, we arrive at a deprojected velocity of $\beta\approx0.043$ and a rest frame energy of $E_\mathrm{rest}\approx6.55$\,keV, which corresponds to Fe\,{\scriptsize XXII}--{\scriptsize XXIII}. 

Assuming a delay of about a week in the appearance of the red-shifted line, we arrive at a distance of $\sim10^{16}$\,cm between the central accretor and the narrow line emitting regions. This distance is difficult to square with a velocity of $0.04c$ in the case that the putative jets were launched during the initial outburst of \maxi, but as we have shown, the $6.3$\,keV line may have gone undetected early in the outburst. While the evidence we have presented may be consistent with emission from an outflow, we are unable to meaningfully distinguish this case with the simpler case of different ionization states. The jet model may be probed by future observations of the source by observatories, such as Chandra, with high spatial resolution and high spectral resolution in the Fe K$\alpha$ complex.

\section{Summary and Conclusions} \label{sec:conclusions}

We have presented MAXI and \nustar\ observations of the low-luminosity transient \maxi, residing in the GC01 cluster. The source was observed twice by \nustar, which was uniquely able to perform observations of the source in outburst due to low angular separation from the Sun. Spectral and timing analyses of the two \nustar\ observations demonstrated a clear change in states from a high soft state to a low hard state. We presented the following results:

\begin{itemize}
    \itemsep0em
    \item MAXI observed a short period of brightening with duration $\sim5$\,days, which then underwent a power law decay for several tens of days.
    \item In the bright state the source reached a luminosity of only $\sim10^{36}\,\mathrm{erg\,cm^{-2}\,s^{-1}}$, corresponding to an Eddington fraction $<0.5\%$.
    \item \nustar\ spectra revealed relativistic disk reflection features. Analysis of these features showed:
    \begin{itemize}
        \itemsep0em
        \item Reflection of thermal emission in the soft state, representing evidence for returning radiation.
        \item An accretor with nearly maximal spin.
        \item An increase in the inner radius of the accretion disk, providing evidence for moderate disk truncation in the low hard state.
        \item Narrow Fe emission components which may differ in energy due to differences in ionization or due to Doppler shifts in an outflow.
    \end{itemize}
    \item Timing analysis of \nustar\ data revealed clear red noise in the hard state. We did not find evidence for features such as QPOs or pulsations, and the continuum noise in both states was too low for us to constrain.
\end{itemize}

Due to the high spin measurement as well as other features such as an anomalously low quiescent luminosity and spectral shapes that resemble those of black hole X-ray binaries, we favor the conclusion that the source is an accreting black hole.

However, the low luminosity of the source is difficult to explain at the same time as evidence for a hot disk which extends very close to the central accretor. We therefore discussed scenarios such as high disk inclination or scattering of emission from the outer disk in order to explain this apparent inconsistency. Further observations of the source during future outbursts -- particularly those observations which can provide measurements of the binary inclination -- will help to elucidate the nature of this intriguing X-ray binary.

\acknowledgments

This work was partially supported under NASA contract No. NNG08FD60C and made use of data from the \nustar\ mission, a project led by the California Institute of Technology, managed by the Jet Propulsion Laboratory, and funded by the National Aeronautics and Space Administration. We thank the \nustar\ Operations, Software, and Calibration teams for support with the execution and analysis of these observations. This research has made use of the \nustar\ Data Analysis Software (NuSTARDAS), jointly developed by the ASI Science Data Center (ASDC, Italy) and the California Institute of Technology (USA). 

R.M.L. acknowledges the support of NASA through Hubble Fellowship Program grant HST-HF2-51440.001. J.H. acknowledges support from an appointment to the NASA Postdoctoral Program at the Goddard Space Flight Center, administered by the Universities Space Research Association under contract with NASA. A.J. would like to acknowledge support from Chandra X-ray Observatory Center's grant GO0-21067X. 

This work also made use of MAXI data provided by RIKEN, JAXA and the MAXI team. Part of this work was financially supported by Grants-in-Aid for Scientific Research 21K03620 (H.N.), 17H06362 (H.N.), and 19K14762 (M.S.) from the Ministry of Education, Culture, Sports, Science and Technology (MEXT) of Japan. 

Finally, we would like to thank the anonymous reviewer whose comments and suggestions significantly improved the quality of this work.

\software{Astropy \citep{Astropy2013,astropy2018}, Corner \citep{corner}, DS9 \citep{Joye2003}, Matplotlib \citep{Hunter:2007}, Scipy \citep{Scipy2020}, Stingray \citep{Huppenkothen2019}, relxill \citep{Dauser2014,Garcia2014}, XSpec \citep{Arnaud1996}}

\bibliography{main}{}

\newcommand{\noop}[1]{}
\begin{thebibliography}{}
\expandafter\ifx\csname natexlab\endcsname\relax\def\natexlab#1{#1}\fi
\providecommand{\url}[1]{\href{#1}{#1}}
\providecommand{\dodoi}[1]{doi:~\href{http://doi.org/#1}{\nolinkurl{#1}}}
\providecommand{\doeprint}[1]{\href{http://ascl.net/#1}{\nolinkurl{http://ascl.net/#1}}}
\providecommand{\doarXiv}[1]{\href{https://arxiv.org/abs/#1}{\nolinkurl{https://arxiv.org/abs/#1}}}

\bibitem[{{Arnaud}(1996)}]{Arnaud1996}
{Arnaud}, K.~A. 1996, in Astronomical Society of the Pacific Conference Series,
  Vol. 101, Astronomical Data Analysis Software and Systems V, ed. G.~H.
  {Jacoby} \& J.~{Barnes}, 17

\bibitem[{{Astropy Collaboration} {et~al.}(2013){Astropy Collaboration},
  {Robitaille}, {Tollerud}, {Greenfield}, {Droettboom}, {Bray}, {Aldcroft},
  {Davis}, {Ginsburg}, {Price-Whelan}, {Kerzendorf}, {Conley}, {Crighton},
  {Barbary}, {Muna}, {Ferguson}, {Grollier}, {Parikh}, {Nair}, {Unther},
  {Deil}, {Woillez}, {Conseil}, {Kramer}, {Turner}, {Singer}, {Fox}, {Weaver},
  {Zabalza}, {Edwards}, {Azalee Bostroem}, {Burke}, {Casey}, {Crawford},
  {Dencheva}, {Ely}, {Jenness}, {Labrie}, {Lim}, {Pierfederici}, {Pontzen},
  {Ptak}, {Refsdal}, {Servillat}, \& {Streicher}}]{Astropy2013}
{Astropy Collaboration}, {Robitaille}, T.~P., {Tollerud}, E.~J., {et~al.} 2013,
  \aap, 558, A33, \dodoi{10.1051/0004-6361/201322068}

\bibitem[{{Astropy Collaboration} {et~al.}(2018){Astropy Collaboration},
  {Price-Whelan}, {Sip{\H{o}}cz}, {G{\"u}nther}, {Lim}, {Crawford}, {Conseil},
  {Shupe}, {Craig}, {Dencheva}, {Ginsburg}, {VanderPlas}, {Bradley},
  {P{\'e}rez-Su{\'a}rez}, {de Val-Borro}, {Aldcroft}, {Cruz}, {Robitaille},
  {Tollerud}, {Ardelean}, {Babej}, {Bach}, {Bachetti}, {Bakanov}, {Bamford},
  {Barentsen}, {Barmby}, {Baumbach}, {Berry}, {Biscani}, {Boquien}, {Bostroem},
  {Bouma}, {Brammer}, {Bray}, {Breytenbach}, {Buddelmeijer}, {Burke},
  {Calderone}, {Cano Rodr{\'\i}guez}, {Cara}, {Cardoso}, {Cheedella}, {Copin},
  {Corrales}, {Crichton}, {D'Avella}, {Deil}, {Depagne}, {Dietrich}, {Donath},
  {Droettboom}, {Earl}, {Erben}, {Fabbro}, {Ferreira}, {Finethy}, {Fox},
  {Garrison}, {Gibbons}, {Goldstein}, {Gommers}, {Greco}, {Greenfield},
  {Groener}, {Grollier}, {Hagen}, {Hirst}, {Homeier}, {Horton}, {Hosseinzadeh},
  {Hu}, {Hunkeler}, {Ivezi{\'c}}, {Jain}, {Jenness}, {Kanarek}, {Kendrew},
  {Kern}, {Kerzendorf}, {Khvalko}, {King}, {Kirkby}, {Kulkarni}, {Kumar},
  {Lee}, {Lenz}, {Littlefair}, {Ma}, {Macleod}, {Mastropietro}, {McCully},
  {Montagnac}, {Morris}, {Mueller}, {Mumford}, {Muna}, {Murphy}, {Nelson},
  {Nguyen}, {Ninan}, {N{\"o}the}, {Ogaz}, {Oh}, {Parejko}, {Parley}, {Pascual},
  {Patil}, {Patil}, {Plunkett}, {Prochaska}, {Rastogi}, {Reddy Janga},
  {Sabater}, {Sakurikar}, {Seifert}, {Sherbert}, {Sherwood-Taylor}, {Shih},
  {Sick}, {Silbiger}, {Singanamalla}, {Singer}, {Sladen}, {Sooley},
  {Sornarajah}, {Streicher}, {Teuben}, {Thomas}, {Tremblay}, {Turner},
  {Terr{\'o}n}, {van Kerkwijk}, {de la Vega}, {Watkins}, {Weaver}, {Whitmore},
  {Woillez}, {Zabalza}, \& {Astropy Contributors}}]{astropy2018}
{Astropy Collaboration}, {Price-Whelan}, A.~M., {Sip{\H{o}}cz}, B.~M., {et~al.}
  2018, \aj, 156, 123, \dodoi{10.3847/1538-3881/aabc4f}

\bibitem[{{Bachetti} {et~al.}(2015){Bachetti}, {Harrison}, {Cook}, {Tomsick},
  {Schmid}, {Grefenstette}, {Barret}, {Boggs}, {Christensen}, {Craig},
  {Fabian}, {F{\"u}rst}, {Gandhi}, {Hailey}, {Kara}, {Maccarone}, {Miller},
  {Pottschmidt}, {Stern}, {Uttley}, {Walton}, {Wilms}, \&
  {Zhang}}]{Bachetti2015}
{Bachetti}, M., {Harrison}, F.~A., {Cook}, R., {et~al.} 2015, \apj, 800, 109,
  \dodoi{10.1088/0004-637X/800/2/109}

\bibitem[{{Bachetti} {et~al.}(2021){Bachetti}, {Markwardt}, {Grefenstette},
  {Gotthelf}, {Kuiper}, {Barret}, {Cook}, {Davis}, {F{\"u}rst}, {Forster},
  {Harrison}, {Madsen}, {Miyasaka}, {Roberts}, {Tomsick}, \&
  {Walton}}]{Bachetti2021}
{Bachetti}, M., {Markwardt}, C.~B., {Grefenstette}, B.~W., {et~al.} 2021, \apj,
  908, 184, \dodoi{10.3847/1538-4357/abd1d6}

\bibitem[{{Belloni} {et~al.}(2002){Belloni}, {Psaltis}, \& {van der
  Klis}}]{Belloni2002}
{Belloni}, T., {Psaltis}, D., \& {van der Klis}, M. 2002, \apj, 572, 392,
  \dodoi{10.1086/340290}

\bibitem[{{Buccheri} {et~al.}(1983){Buccheri}, {Bennett}, {Bignami}, {Bloemen},
  {Boriakoff}, {Caraveo}, {Hermsen}, {Kanbach}, {Manchester}, {Masnou},
  {Mayer-Hasselwander}, {{\"O}zel}, {Paul}, {Sacco}, {Scarsi}, \&
  {Strong}}]{Buccheri1983}
{Buccheri}, R., {Bennett}, K., {Bignami}, G.~F., {et~al.} 1983, \aap, 128, 245

\bibitem[{{Burrows} {et~al.}(2005){Burrows}, {Hill}, {Nousek}, {Kennea},
  {Wells}, {Osborne}, {Abbey}, {Beardmore}, {Mukerjee}, {Short}, {Chincarini},
  {Campana}, {Citterio}, {Moretti}, {Pagani}, {Tagliaferri}, {Giommi},
  {Capalbi}, {Tamburelli}, {Angelini}, {Cusumano}, {Br{\"a}uninger}, {Burkert},
  \& {Hartner}}]{Burrows2005}
{Burrows}, D.~N., {Hill}, J.~E., {Nousek}, J.~A., {et~al.} 2005, \ssr, 120,
  165, \dodoi{10.1007/s11214-005-5097-2}

\bibitem[{{Chakrabarty} {et~al.}(2021){Chakrabarty}, {Jonker}, {Homan}, \& {van
  den Berg}}]{2021ATel14424....1C}
{Chakrabarty}, D., {Jonker}, P.~G., {Homan}, J., \& {van den Berg}, M. 2021,
  The Astronomer's Telegram, 14424, 1

\bibitem[{{Connors} {et~al.}(2019){Connors}, {Garc{\'\i}a}, {Steiner},
  {Grinberg}, {Dauser}, {Sridhar}, {Gatuzz}, {Tomsick}, {Markoff}, \&
  {Harrison}}]{Connors2019}
{Connors}, R. M.~T., {Garc{\'\i}a}, J.~A., {Steiner}, J.~F., {et~al.} 2019,
  \apj, 882, 179, \dodoi{10.3847/1538-4357/ab35df}

\bibitem[{{Connors} {et~al.}(2020){Connors}, {Garc{\'\i}a}, {Dauser},
  {Grinberg}, {Steiner}, {Sridhar}, {Wilms}, {Tomsick}, {Harrison}, \&
  {Licklederer}}]{Connors2020}
{Connors}, R. M.~T., {Garc{\'\i}a}, J.~A., {Dauser}, T., {et~al.} 2020, \apj,
  892, 47, \dodoi{10.3847/1538-4357/ab7afc}

\bibitem[{{Connors} {et~al.}(2021){Connors}, {Garc{\'\i}a}, {Tomsick}, {Hare},
  {Dauser}, {Grinberg}, {Steiner}, {Mastroserio}, {Sridhar}, {Fabian}, {Jiang},
  {Parker}, {Harrison}, \& {Kallman}}]{Connors2021}
{Connors}, R. M.~T., {Garc{\'\i}a}, J.~A., {Tomsick}, J., {et~al.} 2021, \apj,
  909, 146, \dodoi{10.3847/1538-4357/abdd2c}

\bibitem[{{Dauser} {et~al.}(2014){Dauser}, {Garcia}, {Parker}, {Fabian}, \&
  {Wilms}}]{Dauser2014}
{Dauser}, T., {Garcia}, J., {Parker}, M.~L., {Fabian}, A.~C., \& {Wilms}, J.
  2014, \mnras, 444, L100, \dodoi{10.1093/mnrasl/slu125}

\bibitem[{{Davidge} {et~al.}(2016){Davidge}, {Andersen}, {Lardi{\`e}re},
  {Bradley}, {Blain}, {Oya}, {Terada}, {Hayano}, {Lamb}, {Akiyama}, {Ono}, \&
  {Suzuki}}]{Davidge2016}
{Davidge}, T.~J., {Andersen}, D.~R., {Lardi{\`e}re}, O., {et~al.} 2016, \aj,
  152, 173, \dodoi{10.3847/0004-6256/152/6/173}

\bibitem[{{Davies} {et~al.}(2011){Davies}, {Bastian}, {Gieles}, {Seth},
  {Mengel}, \& {Konstantopoulos}}]{Davies2011}
{Davies}, B., {Bastian}, N., {Gieles}, M., {et~al.} 2011, \mnras, 411, 1386,
  \dodoi{10.1111/j.1365-2966.2010.17777.x}

\bibitem[{{Deloye} \& {Bildsten}(2003)}]{Deloye2003}
{Deloye}, C.~J., \& {Bildsten}, L. 2003, \apj, 598, 1217,
  \dodoi{10.1086/379063}

\bibitem[{{Esin} {et~al.}(1997){Esin}, {McClintock}, \& {Narayan}}]{Esin1997}
{Esin}, A.~A., {McClintock}, J.~E., \& {Narayan}, R. 1997, \apj, 489, 865,
  \dodoi{10.1086/304829}

\bibitem[{Foreman-Mackey(2016)}]{corner}
Foreman-Mackey, D. 2016, The Journal of Open Source Software, 1, 24,
  \dodoi{10.21105/joss.00024}

\bibitem[{{Garc{\'\i}a} {et~al.}(2013){Garc{\'\i}a}, {Dauser}, {Reynolds},
  {Kallman}, {McClintock}, {Wilms}, \& {Eikmann}}]{Garcia2013}
{Garc{\'\i}a}, J., {Dauser}, T., {Reynolds}, C.~S., {et~al.} 2013, \apj, 768,
  146, \dodoi{10.1088/0004-637X/768/2/146}

\bibitem[{{Garc{\'\i}a} \& {Kallman}(2010)}]{Garcia2010}
{Garc{\'\i}a}, J., \& {Kallman}, T.~R. 2010, \apj, 718, 695,
  \dodoi{10.1088/0004-637X/718/2/695}

\bibitem[{{Garc{\'\i}a} {et~al.}(2014){Garc{\'\i}a}, {Dauser}, {Lohfink},
  {Kallman}, {Steiner}, {McClintock}, {Brenneman}, {Wilms}, {Eikmann},
  {Reynolds}, \& {Tombesi}}]{Garcia2014}
{Garc{\'\i}a}, J., {Dauser}, T., {Lohfink}, A., {et~al.} 2014, \apj, 782, 76,
  \dodoi{10.1088/0004-637X/782/2/76}

\bibitem[{{Garc{\'\i}a} {et~al.}(2022){Garc{\'\i}a}, {Dauser}, {Ludlam},
  {Parker}, {Fabian}, {Harrison}, \& {Wilms}}]{Garcia2021}
{Garc{\'\i}a}, J.~A., {Dauser}, T., {Ludlam}, R., {et~al.} 2022, \apj, 926, 13,
  \dodoi{10.3847/1538-4357/ac3cb7}

\bibitem[{{Garcia} {et~al.}(2001){Garcia}, {McClintock}, {Narayan}, {Callanan},
  {Barret}, \& {Murray}}]{Garcia2001}
{Garcia}, M.~R., {McClintock}, J.~E., {Narayan}, R., {et~al.} 2001, \apjl, 553,
  L47, \dodoi{10.1086/320494}

\bibitem[{{Gehrels} {et~al.}(2004){Gehrels}, {Chincarini}, {Giommi}, {Mason},
  {Nousek}, {Wells}, {White}, {Barthelmy}, {Burrows}, {Cominsky}, {Hurley},
  {Marshall}, {M{\'e}sz{\'a}ros}, {Roming}, {Angelini}, {Barbier}, {Belloni},
  {Campana}, {Caraveo}, {Chester}, {Citterio}, {Cline}, {Cropper}, {Cummings},
  {Dean}, {Feigelson}, {Fenimore}, {Frail}, {Fruchter}, {Garmire}, {Gendreau},
  {Ghisellini}, {Greiner}, {Hill}, {Hunsberger}, {Krimm}, {Kulkarni}, {Kumar},
  {Lebrun}, {Lloyd-Ronning}, {Markwardt}, {Mattson}, {Mushotzky}, {Norris},
  {Osborne}, {Paczynski}, {Palmer}, {Park}, {Parsons}, {Paul}, {Rees},
  {Reynolds}, {Rhoads}, {Sasseen}, {Schaefer}, {Short}, {Smale}, {Smith},
  {Stella}, {Tagliaferri}, {Takahashi}, {Tashiro}, {Townsley}, {Tueller},
  {Turner}, {Vietri}, {Voges}, {Ward}, {Willingale}, {Zerbi}, \&
  {Zhang}}]{Gehrels2004}
{Gehrels}, N., {Chincarini}, G., {Giommi}, P., {et~al.} 2004, \apj, 611, 1005,
  \dodoi{10.1086/422091}

\bibitem[{{Gendreau} \& {Arzoumanian}(2017)}]{Gendreau2017}
{Gendreau}, K., \& {Arzoumanian}, Z. 2017, Nature Astronomy, 1, 895,
  \dodoi{10.1038/s41550-017-0301-3}

\bibitem[{{Goodman} \& {Weare}(2010)}]{Goodman2010}
{Goodman}, J., \& {Weare}, J. 2010, Communications in Applied Mathematics and
  Computational Science, 5, 65, \dodoi{10.2140/camcos.2010.5.65}

\bibitem[{{Hare} {et~al.}(2018){Hare}, {Kargaltsev}, \& {Rangelov}}]{Hare2018}
{Hare}, J., {Kargaltsev}, O., \& {Rangelov}, B. 2018, \apj, 865, 33,
  \dodoi{10.3847/1538-4357/aad90d}

\bibitem[{{Hare} {et~al.}(2021){Hare}, {Yang}, {Kargaltsev}, {Rangelov},
  {Pike}, \& {Tomsick}}]{Hare2021}
{Hare}, J., {Yang}, H., {Kargaltsev}, O., {et~al.} 2021, The Astronomer's
  Telegram, 14499, 1

\bibitem[{{Harrison} {et~al.}(2013){Harrison}, {Craig}, {Christensen},
  {Hailey}, {Zhang}, {Boggs}, {Stern}, {Cook}, {Forster}, {Giommi},
  {Grefenstette}, {Kim}, {Kitaguchi}, {Koglin}, {Madsen}, {Mao}, {Miyasaka},
  {Mori}, {Perri}, {Pivovaroff}, {Puccetti}, {Rana}, {Westergaard}, {Willis},
  {Zoglauer}, {An}, {Bachetti}, {Barri{\`e}re}, {Bellm}, {Bhalerao},
  {Brejnholt}, {Fuerst}, {Liebe}, {Markwardt}, {Nynka}, {Vogel}, {Walton},
  {Wik}, {Alexander}, {Cominsky}, {Hornschemeier}, {Hornstrup}, {Kaspi},
  {Madejski}, {Matt}, {Molendi}, {Smith}, {Tomsick}, {Ajello}, {Ballantyne},
  {Balokovi{\'c}}, {Barret}, {Bauer}, {Blandford}, {Brandt}, {Brenneman},
  {Chiang}, {Chakrabarty}, {Chenevez}, {Comastri}, {Dufour}, {Elvis}, {Fabian},
  {Farrah}, {Fryer}, {Gotthelf}, {Grindlay}, {Helfand}, {Krivonos}, {Meier},
  {Miller}, {Natalucci}, {Ogle}, {Ofek}, {Ptak}, {Reynolds}, {Rigby},
  {Tagliaferri}, {Thorsett}, {Treister}, \& {Urry}}]{Harrison2013}
{Harrison}, F.~A., {Craig}, W.~W., {Christensen}, F.~E., {et~al.} 2013, \apj,
  770, 103, \dodoi{10.1088/0004-637X/770/2/103}

\bibitem[{{Hogg} \& {Foreman-Mackey}(2018)}]{Hogg2018}
{Hogg}, D.~W., \& {Foreman-Mackey}, D. 2018, \apjs, 236, 11,
  \dodoi{10.3847/1538-4365/aab76e}

\bibitem[{Hunter(2007)}]{Hunter:2007}
Hunter, J.~D. 2007, Computing in Science \& Engineering, 9, 90,
  \dodoi{10.1109/MCSE.2007.55}

\bibitem[{{Huppenkothen} {et~al.}(2019){Huppenkothen}, {Bachetti}, {Stevens},
  {Migliari}, {Balm}, {Hammad}, {Khan}, {Mishra}, {Rashid}, {Sharma}, {Martinez
  Ribeiro}, \& {Valles Blanco}}]{Huppenkothen2019}
{Huppenkothen}, D., {Bachetti}, M., {Stevens}, A.~L., {et~al.} 2019, \apj, 881,
  39, \dodoi{10.3847/1538-4357/ab258d}

\bibitem[{{Joye} \& {Mandel}(2003)}]{Joye2003}
{Joye}, W.~A., \& {Mandel}, E. 2003, in Astronomical Society of the Pacific
  Conference Series, Vol. 295, Astronomical Data Analysis Software and Systems
  XII, ed. H.~E. {Payne}, R.~I. {Jedrzejewski}, \& R.~N. {Hook}, 489

\bibitem[{{Kaastra} \& {Bleeker}(2016)}]{Kaastra2016}
{Kaastra}, J.~S., \& {Bleeker}, J.~A.~M. 2016, \aap, 587, A151,
  \dodoi{10.1051/0004-6361/201527395}

\bibitem[{{Kennea} {et~al.}(2021){Kennea}, {Bahramian}, {Evans}, {Beardmore},
  {Krimm}, {Romano}, {Yamaoka}, {Serino}, \& {Negoro}}]{Kennea2021}
{Kennea}, J.~A., {Bahramian}, A., {Evans}, P.~A., {et~al.} 2021, The
  Astronomer's Telegram, 14420, 1

\bibitem[{{Kobulnicky} {et~al.}(2005){Kobulnicky}, {Monson}, {Buckalew},
  {Darnel}, {Uzpen}, {Meade}, {Babler}, {Indebetouw}, {Whitney}, {Watson},
  {Churchwell}, {Wolfire}, {Wolff}, {Clemens}, {Shah}, {Bania}, {Benjamin},
  {Cohen}, {Dickey}, {Jackson}, {Marston}, {Mathis}, {Mercer}, {Stauffer},
  {Stolovy}, {Norris}, {Kutyrev}, {Canterna}, \& {Pierce}}]{Kobulnicky2005}
{Kobulnicky}, H.~A., {Monson}, A.~J., {Buckalew}, B.~A., {et~al.} 2005, \aj,
  129, 239, \dodoi{10.1086/426337}

\bibitem[{{Koljonen} \& {Tomsick}(2020)}]{Koljonen2020}
{Koljonen}, K.~I.~I., \& {Tomsick}, J.~A. 2020, \aap, 639, A13,
  \dodoi{10.1051/0004-6361/202037882}

\bibitem[{{Kremer} {et~al.}(2018){Kremer}, {Chatterjee}, {Rodriguez}, \&
  {Rasio}}]{Kremer2018}
{Kremer}, K., {Chatterjee}, S., {Rodriguez}, C.~L., \& {Rasio}, F.~A. 2018,
  \apj, 852, 29, \dodoi{10.3847/1538-4357/aa99df}

\bibitem[{{Kubota} \& {Makishima}(2004)}]{Kubota2004}
{Kubota}, A., \& {Makishima}, K. 2004, \apj, 601, 428, \dodoi{10.1086/380433}

\bibitem[{{Kubota} {et~al.}(1998){Kubota}, {Tanaka}, {Makishima}, {Ueda},
  {Dotani}, {Inoue}, \& {Yamaoka}}]{Kubota1998}
{Kubota}, A., {Tanaka}, Y., {Makishima}, K., {et~al.} 1998, \pasj, 50, 667,
  \dodoi{10.1093/pasj/50.6.667}

\bibitem[{{Lasota}(2001)}]{Lasota2001}
{Lasota}, J.-P. 2001, \nar, 45, 449, \dodoi{10.1016/S1387-6473(01)00112-9}

\bibitem[{{Lazar} {et~al.}(2021){Lazar}, {Tomsick}, {Pike}, {Bachetti},
  {Buisson}, {Connors}, {Fabian}, {Fuerst}, {Garc{\'\i}a}, {Hare}, {Jiang},
  {Shaw}, \& {Walton}}]{Lazar2021}
{Lazar}, H., {Tomsick}, J.~A., {Pike}, S.~N., {et~al.} 2021, \apj, 921, 155,
  \dodoi{10.3847/1538-4357/ac1bab}

\bibitem[{{Maccarone}(2003)}]{Maccarone2003}
{Maccarone}, T.~J. 2003, \aap, 409, 697, \dodoi{10.1051/0004-6361:20031146}

\bibitem[{{Maitra} \& {Bailyn}(2006)}]{Maitra2006}
{Maitra}, D., \& {Bailyn}, C.~D. 2006, \apj, 637, 992, \dodoi{10.1086/498422}

\bibitem[{{Matsuoka} {et~al.}(2009){Matsuoka}, {Kawasaki}, {Ueno}, {Tomida},
  {Kohama}, {Suzuki}, {Adachi}, {Ishikawa}, {Mihara}, {Sugizaki}, {Isobe},
  {Nakagawa}, {Tsunemi}, {Miyata}, {Kawai}, {Kataoka}, {Morii}, {Yoshida},
  {Negoro}, {Nakajima}, {Ueda}, {Chujo}, {Yamaoka}, {Yamazaki}, {Nakahira},
  {You}, {Ishiwata}, {Miyoshi}, {Eguchi}, {Hiroi}, {Katayama}, \&
  {Ebisawa}}]{Matsuoka2009}
{Matsuoka}, M., {Kawasaki}, K., {Ueno}, S., {et~al.} 2009, \pasj, 61, 999,
  \dodoi{10.1093/pasj/61.5.999}

\bibitem[{{McClintock} \& {Remillard}(2006)}]{McClintock2006}
{McClintock}, J.~E., \& {Remillard}, R.~A. 2006, {Black hole binaries},
  Vol.~39, 157--213

\bibitem[{{Migliari} {et~al.}(2002){Migliari}, {Fender}, \&
  {M{\'e}ndez}}]{Migliari2002}
{Migliari}, S., {Fender}, R., \& {M{\'e}ndez}, M. 2002, Science, 297, 1673,
  \dodoi{10.1126/science.1073660}

\bibitem[{{Mihara} {et~al.}(2011){Mihara}, {Nakajima}, {Sugizaki}, {Serino},
  {Matsuoka}, {Kohama}, {Kawasaki}, {Tomida}, {Ueno}, {Kawai}, {Kataoka},
  {Morii}, {Yoshida}, {Yamaoka}, {Nakahira}, {Negoro}, {Isobe}, {Yamauchi}, \&
  {Sakurai}}]{Mihara2011}
{Mihara}, T., {Nakajima}, M., {Sugizaki}, M., {et~al.} 2011, \pasj, 63, S623,
  \dodoi{10.1093/pasj/63.sp3.S623}

\bibitem[{{Mihara} {et~al.}(2021){Mihara}, {Negoro}, {Shidatsu}, {Sugizaki},
  {Serino}, {Pike}, {Miyasaka}, {Nakajima}, {Aoki}, {Takagi}, {Kobayashi},
  {Asakura}, {Seino}, {Tamagawa}, {Matsuoka}, {Sakamoto}, {Sugita}, {Nishida},
  {Komachi}, {Yoshida}, {Tsuboi}, {Iwakiri}, {Sasaki}, {Kawai}, {Okamoto},
  {Kitakoga}, {Kawai}, {Adachi}, {Niwano}, {Hosokawa}, {Nakahira}, {Sugawara},
  {Ueno}, {Tomida}, {Ishikawa}, {Tominaga}, {Nagatsuka}, {Ueda}, {Yamada},
  {Ogawa}, {Setoguchi}, {Yoshitake}, {Goto}, {Uematsu}, {Tsunemi}, {Yamauchi},
  {Kurogi}, {Miike}, {Kawamuro}, {Yamaoka}, \& {Kawakub}}]{Mihara2021}
{Mihara}, T., {Negoro}, H., {Shidatsu}, M., {et~al.} 2021, The Astronomer's
  Telegram, 14327, 1

\bibitem[{{Miller} {et~al.}(2002){Miller}, {Fabian}, {in't Zand}, {Reynolds},
  {Wijnands}, {Nowak}, \& {Lewin}}]{Miller2002}
{Miller}, J.~M., {Fabian}, A.~C., {in't Zand}, J.~J.~M., {et~al.} 2002, \apjl,
  577, L15, \dodoi{10.1086/344047}

\bibitem[{{Miller} {et~al.}(2021){Miller}, {Sanna}, {Burderi}, {Di Salvo},
  {Chakrabarty}, {Ng}, \& {Gendreau}}]{Miller2021}
{Miller}, J.~M., {Sanna}, A., {Burderi}, L., {et~al.} 2021, The Astronomer's
  Telegram, 14429, 1

\bibitem[{{Morii} {et~al.}(2010){Morii}, {Kawai}, {Sugimori}, {Suzuki},
  {Negoro}, {Sugizaki}, {Nakajima}, {Mihara}, \& {Matsuoka}}]{Morii2010}
{Morii}, M., {Kawai}, N., {Sugimori}, K., {et~al.} 2010, in American Institute
  of Physics Conference Series, Vol. 1279, Deciphering the Ancient Universe
  with Gamma-ray Bursts, ed. N.~{Kawai} \& S.~{Nagataki}, 391--393,
  \dodoi{10.1063/1.3509322}

\bibitem[{{Negoro} {et~al.}(2016){Negoro}, {Kohama}, {Serino}, {Saito},
  {Takahashi}, {Miyoshi}, {Ozawa}, {Suwa}, {Asada}, {Fukushima}, {Eguchi},
  {Kawai}, {Kennea}, {Mihara}, {Morii}, {Nakahira}, {Ogawa}, {Sugawara},
  {Tomida}, {Ueno}, {Ishikawa}, {Isobe}, {Kawamuro}, {Kimura}, {Masumitsu},
  {Nakagawa}, {Nakajima}, {Sakamoto}, {Shidatsu}, {Sugizaki}, {Sugimoto},
  {Suzuki}, {Takagi}, {Tanaka}, {Tsuboi}, {Tsunemi}, {Ueda}, {Yamaoka},
  {Yamauchi}, {Yoshida}, \& {Matsuoka}}]{Negoro2016}
{Negoro}, H., {Kohama}, M., {Serino}, M., {et~al.} 2016, \pasj, 68, S1,
  \dodoi{10.1093/pasj/psw016}

\bibitem[{{Negoro} {et~al.}(2020){Negoro}, {Mihara}, {Seino}, {Nakajima},
  {Aoki}, {Kobayashi}, {Takagi}, {Asakura}, {Tamagawa}, {Matsuoka}, {Sakamoto},
  {Serino}, {Sugita}, {Nishida}, {Komachi}, {Yoshida}, {Tsuboi}, {Iwakiri},
  {Sasaki}, {Kawai}, {Okamoto}, {Kitakoga}, {Shidatsu}, {Kawai}, {Adachi},
  {Niwano}, {Hosokawa}, {Nakahira}, {Sugawara}, {Ueno}, {Tomida}, {Ishikawa},
  {Tominaga}, {Nagatsuka}, {Ueda}, {Yamada}, {Ogawa}, {Setoguchi}, {Yoshitake},
  {Goto}, {Uematsu}, {Tsunemi}, {Yamauchi}, {Kurogi}, {Miike}, {Kawamuro},
  {Yamaoka}, {Kawakubo}, {Sugizaki}, \& {MAXI Team}}]{Negoro2020}
{Negoro}, H., {Mihara}, T., {Seino}, K., {et~al.} 2020, The Astronomer's
  Telegram, 14292, 1

\bibitem[{{Pike} {et~al.}(2020){Pike}, {Harrison}, {Forster}, {Grefenstette},
  {Miyasaka}, {Xu}, \& {Tomsick}}]{Pike2020}
{Pike}, S.~N., {Harrison}, F.~A., {Forster}, K., {et~al.} 2020, The
  Astronomer's Telegram, 14290, 1

\bibitem[{{Revnivtsev} {et~al.}(2002){Revnivtsev}, {Gilfanov}, {Churazov}, \&
  {Sunyaev}}]{Revnivtsev2002}
{Revnivtsev}, M., {Gilfanov}, M., {Churazov}, E., \& {Sunyaev}, R. 2002, \aap,
  391, 1013, \dodoi{10.1051/0004-6361:20020865}

\bibitem[{{Shakura} \& {Sunyaev}(1973)}]{Shakura1973}
{Shakura}, N.~I., \& {Sunyaev}, R.~A. 1973, \aap, 500, 33

\bibitem[{{Shimura} \& {Takahara}(1995)}]{Shimura1995}
{Shimura}, T., \& {Takahara}, F. 1995, \apj, 445, 780, \dodoi{10.1086/175740}

\bibitem[{{Sugizaki} {et~al.}(2001){Sugizaki}, {Mitsuda}, {Kaneda},
  {Matsuzaki}, {Yamauchi}, \& {Koyama}}]{Sugizaki2001}
{Sugizaki}, M., {Mitsuda}, K., {Kaneda}, H., {et~al.} 2001, \apjs, 134, 77,
  \dodoi{10.1086/320358}

\bibitem[{{Sugizaki} {et~al.}(2011){Sugizaki}, {Mihara}, {Serino}, {Yamamoto},
  {Matsuoka}, {Kohama}, {Tomida}, {Ueno}, {Kawai}, {Morii}, {Sugimori},
  {Nakahira}, {Yamaoka}, {Yoshida}, {Nakajima}, {Negoro}, {Eguchi}, {Isobe},
  {Ueda}, \& {Tsunemi}}]{Sugizaki2011}
{Sugizaki}, M., {Mihara}, T., {Serino}, M., {et~al.} 2011, \pasj, 63, S635,
  \dodoi{10.1093/pasj/63.sp3.S635}

\bibitem[{{Takagi} {et~al.}(2020){Takagi}, {Negoro}, {Serino}, {Nakajima},
  {Aoki}, {Kobayashi}, {Asakura}, {Seino}, {Mihara}, {Tamagawa}, {Matsuoka},
  {Sakamoto}, {Sugita}, {Nishida}, {Komachi}, {Yoshida}, {Tsuboi}, {Iwakiri},
  {Sasaki}, {Kawai}, {Okamoto}, {Kitakoga}, {Shidatsu}, {Kawai}, {Adachi},
  {Niwano}, {Hosokawa}, {Nakahira}, {Sugawara}, {Ueno}, {Tomida}, {Ishikawa},
  {Tominaga}, {Nagatsuka}, {Ueda}, {Yamada}, {Ogawa}, {Setoguchi}, {Yoshitake},
  {Goto}, {Uematsu}, {Tsunemi}, {Yamauchi}, {Kurogi}, {Miike}, {Kawamuro},
  {Yamaoka}, {Kawakubo}, {Sugizaki}, \& {MAXI Team}}]{Takagi2020}
{Takagi}, R., {Negoro}, H., {Serino}, M., {et~al.} 2020, The Astronomer's
  Telegram, 14282, 1

\bibitem[{{Tremou} {et~al.}(2021){Tremou}, {Carotenuto}, {Fender}, {Woudt},
  {Miller-Jones}, {Motta}, \& {ThunderKAT Collaboration}}]{Tremou2021}
{Tremou}, E., {Carotenuto}, F., {Fender}, R., {et~al.} 2021, The Astronomer's
  Telegram, 14432, 1

\bibitem[{{Tsygankov} {et~al.}(2017){Tsygankov}, {Wijnands}, {Lutovinov},
  {Degenaar}, \& {Poutanen}}]{Tsygankov2017}
{Tsygankov}, S.~S., {Wijnands}, R., {Lutovinov}, A.~A., {Degenaar}, N., \&
  {Poutanen}, J. 2017, \mnras, 470, 126, \dodoi{10.1093/mnras/stx1255}

\bibitem[{{Vahdat Motlagh} {et~al.}(2019){Vahdat Motlagh}, {Kalemci}, \&
  {Maccarone}}]{Motlagh2019}
{Vahdat Motlagh}, A., {Kalemci}, E., \& {Maccarone}, T.~J. 2019, \mnras, 485,
  2744, \dodoi{10.1093/mnras/stz569}

\bibitem[{{Verner} {et~al.}(1996){Verner}, {Ferland}, {Korista}, \&
  {Yakovlev}}]{Verner1996}
{Verner}, D.~A., {Ferland}, G.~J., {Korista}, K.~T., \& {Yakovlev}, D.~G. 1996,
  \apj, 465, 487, \dodoi{10.1086/177435}

\bibitem[{{Virtanen} {et~al.}(2020){Virtanen}, {Gommers}, {Oliphant},
  {Haberland}, {Reddy}, {Cournapeau}, {Burovski}, {Peterson}, {Weckesser},
  {Bright}, {van der Walt}, {Brett}, {Wilson}, {Millman}, {Mayorov}, {Nelson},
  {Jones}, {Kern}, {Larson}, {Carey}, {Polat}, {Feng}, {Moore}, {VanderPlas},
  {Laxalde}, {Perktold}, {Cimrman}, {Henriksen}, {Quintero}, {Harris},
  {Archibald}, {Ribeiro}, {Pedregosa}, {van Mulbregt}, \& {SciPy 1. 0
  Contributors}}]{Scipy2020}
{Virtanen}, P., {Gommers}, R., {Oliphant}, T.~E., {et~al.} 2020, Nature
  Methods, 17, 261, \dodoi{10.1038/s41592-019-0686-2}

\bibitem[{{Wachter} {et~al.}(1979){Wachter}, {Leach}, \&
  {Kellogg}}]{Wachter1979}
{Wachter}, K., {Leach}, R., \& {Kellogg}, E. 1979, \apj, 230, 274,
  \dodoi{10.1086/157084}

\bibitem[{{Weisskopf} {et~al.}(2000){Weisskopf}, {Tananbaum}, {Van Speybroeck},
  \& {O'Dell}}]{Weisskopf2000}
{Weisskopf}, M.~C., {Tananbaum}, H.~D., {Van Speybroeck}, L.~P., \& {O'Dell},
  S.~L. 2000, in Society of Photo-Optical Instrumentation Engineers (SPIE)
  Conference Series, Vol. 4012, X-Ray Optics, Instruments, and Missions III,
  ed. J.~E. {Truemper} \& B.~{Aschenbach}, 2--16, \dodoi{10.1117/12.391545}

\bibitem[{{Wijnands} \& {van der Klis}(2000)}]{Wijnands2000}
{Wijnands}, R., \& {van der Klis}, M. 2000, \apjl, 528, L93,
  \dodoi{10.1086/312439}

\bibitem[{{Wilms} {et~al.}(2000){Wilms}, {Allen}, \& {McCray}}]{Wilms2000}
{Wilms}, J., {Allen}, A., \& {McCray}, R. 2000, \apj, 542, 914,
  \dodoi{10.1086/317016}

\bibitem[{{Zdziarski} {et~al.}(1996){Zdziarski}, {Johnson}, \&
  {Magdziarz}}]{Zdziarski1996}
{Zdziarski}, A.~A., {Johnson}, W.~N., \& {Magdziarz}, P. 1996, \mnras, 283,
  193, \dodoi{10.1093/mnras/283.1.193}

\bibitem[{{{\.Z}ycki} {et~al.}(1999){{\.Z}ycki}, {Done}, \&
  {Smith}}]{Zycki1999}
{{\.Z}ycki}, P.~T., {Done}, C., \& {Smith}, D.~A. 1999, \mnras, 309, 561,
  \dodoi{10.1046/j.1365-8711.1999.02885.x}

\end{thebibliography}

\bibliographystyle{aasjournal}

\appendix

\renewcommand\thefigure{\thesection.\arabic{figure}}
\setcounter{figure}{0}
\section{Breaking the Spectral fit degeneracy} \label{sec:individualfits}

\begin{figure*}[h]
\begin{center}
\includegraphics[width=0.60\textwidth]{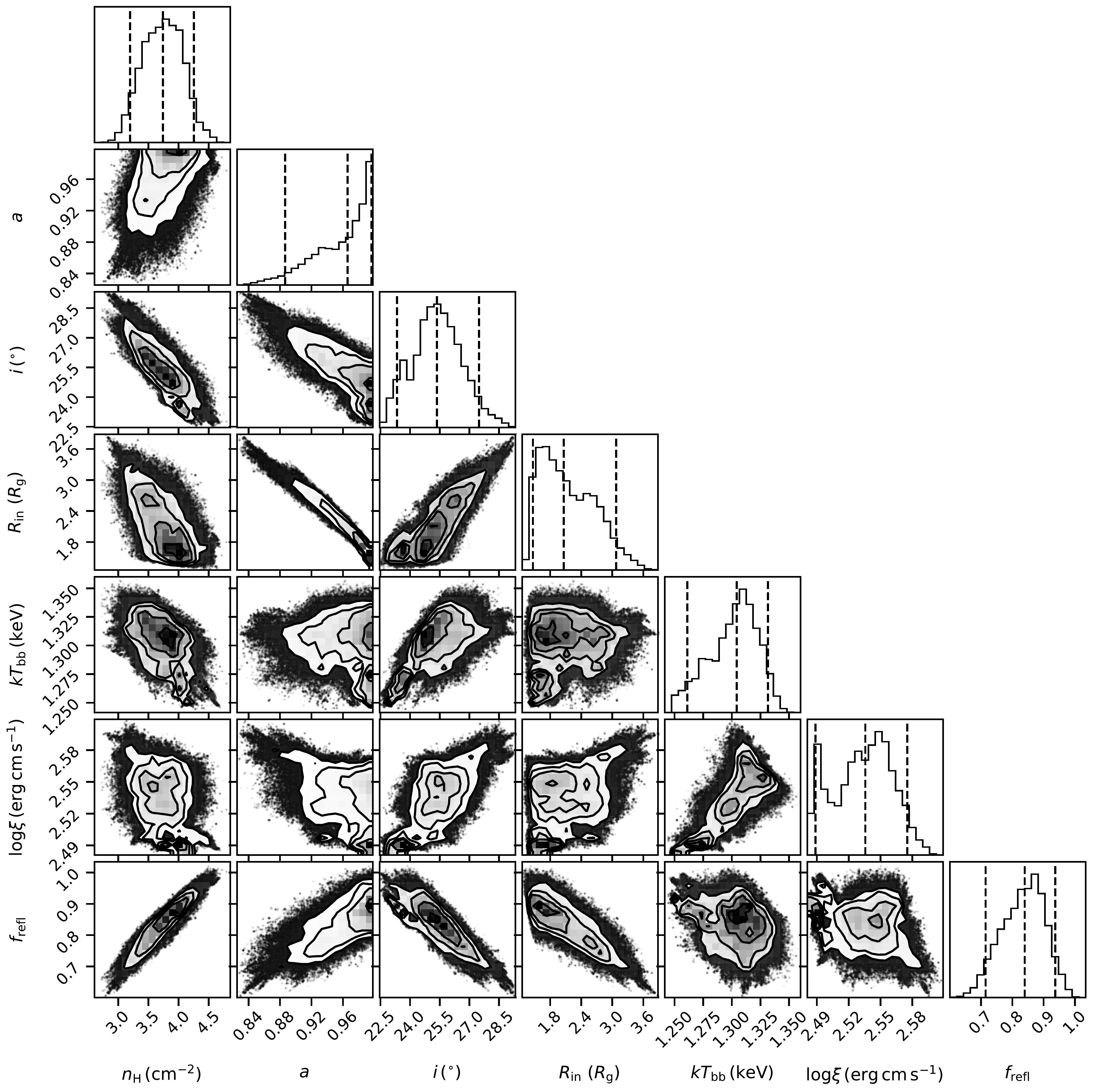}
\vfill
\includegraphics[width=0.60\textwidth]{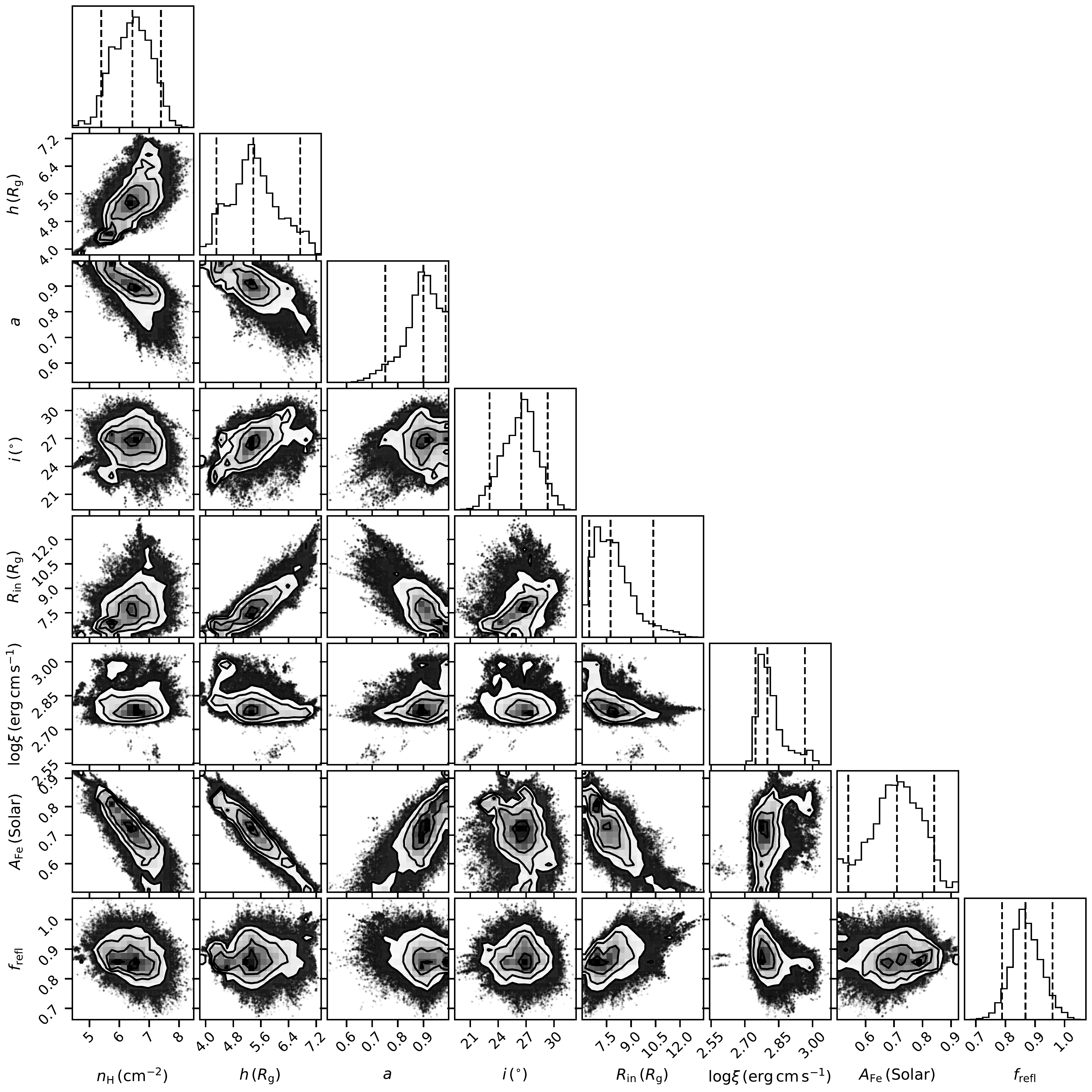}
\caption{Distributions of selected parameters resulting from MCMC analysis of each spectrum individually, without joint fitting. The dashed lines shown in the one-dimensional histograms represent, from left to right, the $5\%$, $50\%$, and $95\%$ quantiles, while the contours shown in the two-dimensional histograms represent confidence regions of increasing sigma. Top: soft state parameters. Bottom: hard state parameters.
\label{fig:individual_corner}}
\end{center}
\end{figure*}

Fitting the spectra with their respective best fit models (described in Section \ref{sec:spectra}) individually, rather than performing a joint fit, resulted in significant degeneracies between a number of parameters. In Figure \ref{fig:individual_corner} we show the resulting distributions for those parameters which demonstrate this degeneracy. In particular, although the spin is consistent with values above $a=0.7$, it is not particularly well-constrained. 

In the soft state, the spin and the inclination show a clear dependence on one another, as do the spin and the inner radius. The latter degeneracy is due to the proximity of the inner radius with the inner-most stable circular orbit, the physical value of which depends directly on the spin of the central accretor. The strength of reflection, parametrized by the reflection fraction, $f_\mathrm{refl}$, also shows a clear correlation with a number of parameters, including the absorption column density, the spin, the inclination, and the inner disk radius. In the hard state, on the other hand, the spin appears to be controlled mainly by the iron abundance, $A_\mathrm{Fe}$, the absorption column density, the height of the lamppost, and the inner disk radius, and it shows less dependence on the inclination of the inner disk. In the soft state, the iron abundance consistently tended towards the lower limit of the model, $A_\mathrm{Fe}=0.5\,A_\mathrm{Fe,\odot}$, so we froze the parameter at this value.

\begin{figure*}
\begin{center}
\includegraphics[width=0.98\textwidth]{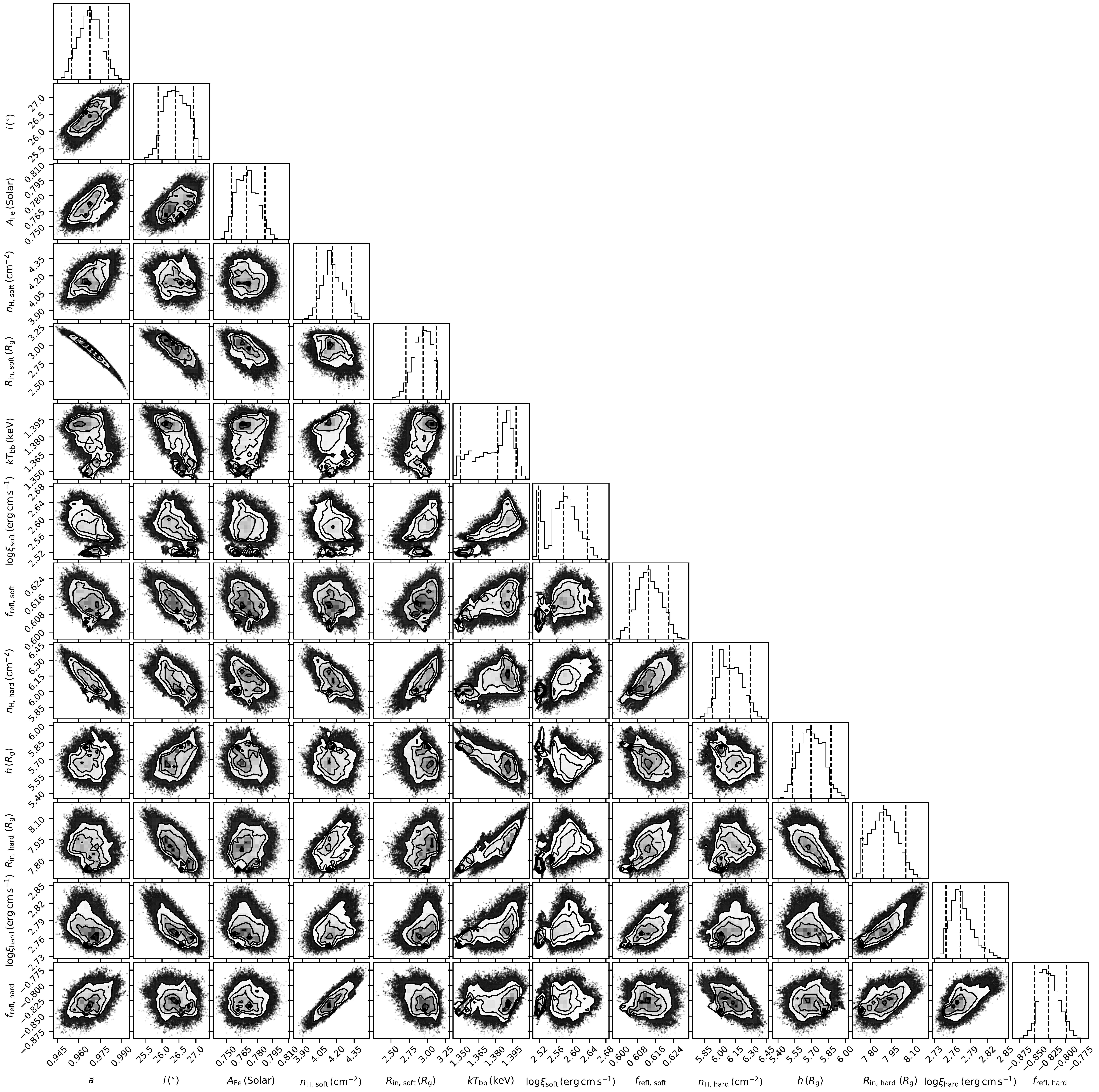}
\caption{Distributions of selected parameters resulting from MCMC analysis of the NuSTAR spectra with joint fitting where the parameters $a$, $i$, and $A_\mathrm{Fe}$ were tied between the soft and hard states. The dashed lines shown in the one-dimensional histograms represent, from left to right, the $5\%$, $50\%$, and $95\%$ quantiles, while the contours shown in the two-dimensional histograms represent confidence regions of increasing sigma.
\label{fig:joint_corner}}
\end{center}
\end{figure*}

We also show the parameter distributions for the joint fits in Figure \ref{fig:joint_corner} in order to demonstrate the superior constraints we were able to achieve via this method. Importantly, the spin and the inner disk inclination were constrained very well compared to the individual fits. A few distinct correlations remain, with the spin depending on the inner disk radius of the soft state and the lamppost height during the hard state, and the inclination being controlled by the ionization of the disk in the hard state. Both values also show a noticeable dependence on the absorption column density. Notably, using this joint fitting method, we obtained a well-constrained distribution for the iron abundance, which we were not able to achieve via the individual fits.

\end{document}